\documentclass[12pt,aps]{revtex4}
\usepackage{epsf}
\usepackage{amsmath,amsthm,amssymb}
\usepackage{color}
\usepackage{dcolumn}
\usepackage{bm}
\usepackage{graphicx}
\usepackage{graphicx,epsfig}
\usepackage{esint}

\begin{document}

\author{Zhong-Can Ou-Yang$^{1}$\thanks{%
{} E-mail address: oy@itp.ac.cn, }, Tao Xu$^{1,2}$\thanks{%
{} E-mail address: xutao@hust.edu.cn, }}
\title{Liquid Crystal Theory of Biomembranes }
\affiliation{\\
$^{1}$Institute of theoretical physics, Chinese Academy of Science, P.O. Box 2735,
Beijing 100080, China\\
$^{2}$Fusion Center, College of Electrical and Electronic Engineering,
Huazhong University of Science and Technology, Wuhan 430074, PRC.}

\begin{abstract}
Biomembranes, primarily composed of lipid bilayers, are not merely passive barriers but dynamic and complex materials whose shapes are governed by the principles of soft matter physics. This review explores the shape problem in biomembranes through the lens of material science and liquid crystal theory.
Beginning with classical analogies to crystals and soap bubbles, it details the application of the Helfrich elastic model to explain the biconcave shape of red blood cells. The discussion extends to multi-layer systems, drawing parallels between the focal conic structures of smectic liquid crystals, the
geometries of fullerenes and carbon nanotubes, and the reversible
transitions in peptide assemblies. Furthermore, it examines icosahedral self-assemblies and shape formation in two-dimensional lipid monolayers at
air/water interfaces. At the end of the paper, we find that the shapes such as cylinders, spheres, tori, biconcave discoids and Delaunay surfaces form a group. This result is merely an intrinsic geometric feature of these shapes and is independent of the biomembrane equation. When the pressure on the
membrane, surface tension, and bending modules meet certain conditions, the biomembrane will take on these shapes. The review concludes by highlighting the unifying power of continuum elastic theories in describing a vast array of membrane morphologies across biological and synthetic systems.

\end{abstract}

\maketitle

%

\section{Introduction}

In 1922, the French mineralogist Georges Friedel, with his son, systematically
laid the experimental foundations \cite{1} of liquid crystal science. His
core contributions were threefold: First, he established the fundamental
phase classification, defining the layered "smectic" and the orientationally
ordered "nematic" phases. Second, using polarizing microscopy, he provided
the first detailed description of the complex focal conic texture in
smectics, recognizing it as an intrinsic feature of the layered structure.
Third, although he did not use the term "Dupin cyclides," his precise description
of the focal conics (an ellipse and a confocal hyperbola) provided the
essential experimental groundwork for all future geometric and theoretical
explanations.

Building directly upon Friedel's observations, William Bragg, in his seminal
paper "The Focal Conic Structure in Smectic Liquid Crystals" \cite{1A},
provided the critical geometric insight. He demonstrated that the seemingly
complex focal conic texture could be described elegantly by a family of
surfaces maintaining constant interlayer spacing. Bragg proved that the
mathematical solution to this condition is a family of surfaces known as
Dupin cyclides, explicitly identifying them as the geometric essence of the
smectic layer arrangement. This work decisively answered the "how" by
connecting the microscopic defects to a precise, universal geometric model.

While Bragg explained the geometry, the fundamental "why"---the underlying
energy minimization principle---was later explained using continuum
elasticity theory. The crucial framework for this was established by
Wolfgang Helfrich in 1973. In his paper "Elastic properties of lipid
bilayers: theory and possible experiments" \cite{1B}, he formulated a
general curvature elastic energy for fluid membranes. This Helfrich free
energy, expressed in terms of mean and Gaussian curvature, created a unified
continuum theory for all thin film systems governed by curvature elasticity.
This model was directly applied to smectic liquid crystals, proving that the
Dupin cyclide configuration described by Bragg is indeed the
energy-minimizing solution under the constraint of layer incompressibility,
thereby completing the theoretical picture from phenomenon to geometry to
energy.

The collaboration between Ou-Yang Zhong-Can and Wolfgang Helfrich in the
late 1980s produced two foundational papers that bridged the gap between the
theory of membrane elasticity and the prediction of complex biological
shapes. Their key publications are \textquotedblleft Instability and
deformation of a spherical vesicle by pressure\textquotedblright \cite{1C},
\textquotedblleft Bending energy of vesicle membranes: General expressions
for the first, second, and third variation of the shape energy and
applications to spheres and cylinders\textquotedblright \cite{1D}.

Building upon Helfrich's curvature elasticity model for membranes, their
rigorous application of variational methods led to the derivation of a
universal differential equation governing vesicle equilibrium shapes. This
pivotal result, known as the \textquotedblleft Zhong-Can-Helfrich
equation \textquotedblright \ provided important quantitative theoretical
framework for explaining intricate biomembrane morphologies, most notably
the biconcave disk shape of red blood cells.

For more than a century, the unique biconcave disk shape of the human red blood
cell has been a major puzzle in biophysics. The analysis is based on the
Helfrich theory of fluid membranes. The lipid bilayer is treated as a
two-dimensional liquid crystal sheet, with its equilibrium shape determined
by minimizing the bending elastic energy. The primary contribution of Naito, Okuda and Ou-Yang
 \cite{1E} was to demonstrate that the condition of zero osmotic pressure difference, the equation of the axisymmetric shape admits a
specific analytic solution that perfectly describes the classic biconcave
disk profile. The work was also profoundly predictive, forecasting novel
shapes such as toroidal vesicles that were subsequently verified by experiment.

In 1992, Prof. Podgornik \cite{1P1} collaborated with Parsegian to establish the theoretical framework describing the fluctuations of fluid membranes in confined spaces. In 2015 \cite{1P2}, he employed osmotic stress small-angle X-ray scattering techniques to precisely measure the bending moduli of different domains within lipid membranes, revealing that the dominant repulsive force between membranes originates from membrane undulations—termed undulatory repulsion.  Prof. Podgornik revealed that the shape of a membrane is not static; its intrinsic vibrations, or fluctuations, couple with electrostatic interactions, van der Waals forces, and other effects to generate new, effective interaction forces. 

Conventional models often treat biological membranes as surfaces bearing fixed charges. However, Prof. Podgornik recognized that the dissociation state of ionizable groups on the membrane—such as amino acid residues—varies dynamically in response to environmental conditions like pH and salt concentration. This mechanism, known as "charge regulation," is crucial for understanding realistic biological environments. In 2019 \cite{1P3}, he demonstrated that in stacked membrane systems, charge regulation fundamentally alters the electrostatic forces between membranes, exhibiting a dependence on pH that differs entirely from the classical boundary conditions of either constant charge or constant potential. Subsequently, in 2022 \cite{1P4}, he uncovered the coupling effect between membrane curvature and charge regulation, showing that a symmetric lipid bilayer can undergo spontaneous charge symmetry breaking due to charge regulation and exhibit anomalous curvature dependence of the free energy as well as nonlinear flexoelectric effects.  Prof. Podgornik and Prof. Andelman et.al. \cite{1P5} established charge regulation as a generalized boundary condition for membrane shape.

Electrostatic repulsion between charged particles in colloidal crystals is the foundation of the stability of the system. However, traditional theories often neglect the influence of lattice elastic deformation on electrostatic interactions. Recently, Prof. Ou-Yang, Prof. Podgornik, and Wu Hao developed a continuum theory \cite{1k} \cite{1m} to systematically explore the coupling effects between the two. They found that under appropriate conditions, mobile charges of the same sign, which normally repel each other, can experience a net effective attraction due to the ‘bridging’ effect of the elastic lattice.

Beyond biophysics, Ou-Yang later expanded his research to elucidate the
complex forms found in diverse soft matter systems, including focal conic
domains in liquid crystals, the structure of carbon nanotubes, and the
assembly of viral capsids.

Prof. Ou-Yang et al. \cite{1G}, \cite{2} have developed the Helfrich theory in
fullerenes and carbon nanotubes. They take the continuum limit of Lenosky's
discrete carbon network model to formulate a curvature elasticity theory in
the same spirit as the Helfrich model in its mathematical form. Both embody
the physical idea of using continuous curvature to describe microscopic
interactions.

The morphology of nanofiber membranes or vesicles formed by the
self-assembly of peptide amphiphiles is similarly governed by a balance
between curvature elastic energy (driven by the molecules' intrinsic
spontaneous curvature) and surface energy. The Helfrich model \cite{1H}
serves as the natural theoretical tool for analyzing their stability.

During the past century, numerous scientists have made outstanding
contributions in these fields. This article does not exhaustively list all of them. This paper is organized as follows. Section $2$ presents shape problem
in material science: crystals and soap bubble, Section $3$ presents shape of
red blood cell and elastic theory of membranes in liquid crystal phases:
Helfrich Model for bilayer vesicles.Section $4$ presents Helfrich model for
multi-layer vesicles (I): focal conic structure of smectic liquid crystal,
Section $5$ presents Helfrich model for multi-layer Vesicle (II): the shapes
of fullerenes and carbon nanotubes, Section $6$ presents Helfrich model for
reversible transition between peptide nanotobes and spheric vesicles induced
by concentrating solution, Section $7$ presents Helfrich model for
icosahedral self-assemblies, Section $8$ presents shape formation in 2D
lipid monolayer at air/water interface, Section $9$ presents structure of
membrane shape equation, and Section $10$ is conclusion.

\section{Membrane shape problem in material science}

The development of membrane shape equations is a fascinating journey through
scientific history, connecting early observations of nature to modern
material science. Here is an introduction to the key develpopmet of membrane
shape research.

N. Stensen made a seminal observation in crystallography: the angles between
corresponding faces of quartz crystals are constant, regardless of their
size or gross shape. Stensen's Law of constants of interfacial angstroms
implied that the external form of a crystal is a direct manifestation of its
internal, periodic atomic structure. This established a crucial
philosophical and scientific link between macroscopic morphology and crystal
symmetry, a concept that would later underpin all theories of equilibrium
shape, including those for membranes.

Building on thermodynamics, G. Wulff established a geometric construction to
determine the equilibrium shape of a crystal or droplet \cite{3}. Wulff's
theorem states that the equilibrium form minimizes the total surface energy
for a given volume. In this shape, the distance from a crystal face to the
center is proportional to its specific surface energy.

A crystal is anisotropic, so its surface free energy per unit area, $\gamma $%
, is not a constant but depends strongly on the crystallographic orientation 
$\overline{n}$ of the surface. i.e.

\begin{equation}
F=\oint \gamma (\overrightarrow{n})dA+\lambda \int dA \label{1}
\end{equation}%
Here $\lambda $ is Lagrange multiplier for a constant volume.

The equilibrium shape of a crystal (or a coherent domain within a material)
at constant temperature and pressure is the one that minimizes the total
surface free energy for a given enclosed volume. Then

\begin{equation}
\delta F=0.  \label{2}
\end{equation}%
This local mechanical equilibrium lead to the Wulff condition

\begin{equation}
\gamma (n)=\lambda \overrightarrow{r}\cdot \overrightarrow{n},  \label{3}
\end{equation}%
where $\overrightarrow{r}\cdot \overrightarrow{n}$ is the perpendicular
distance from the Wulff point to the tangent plane of the surface in
direction $\overrightarrow{n}$.

Through elegant experiments with soap films, J. Plateau studied minimal
surfaces. He demonstrated that a soap film bounded by a wire loop adjusts to
find the shape with the minimum possible surface area. This principle is
mathematically expressed as the condition of zero mean curvature ($H=0$).
His work defined the famous "Plateau's problem" in mathematics.

A soap film's energy is proportional to its surface area due to constant
tension. Therefore, a stable, equilibrium film adopts the shape that
minimizes its total surface area $A$ for a given boundary.

\begin{equation}
F=\oint dA.  \label{4}
\end{equation}%
The result of the minimization $\delta F=0$ is the condition that the mean
curvature must be zero everywhere on the surface \cite{4}%
\begin{equation}
H=0.  \label{5}
\end{equation}%
A surface with zero mean curvature is known as a minimal surface. This is
the mathematical definition of the shapes Plateau observed physically.

Studying capillary action, T. Young and P.S. Laplace (1805-1806) founded the
theory of capillary surfaces. The relevant free energy for a fluid interface
at constant temperature is the Helmholtz free energy, F. For a system with a
constant surface tension $\gamma $, the free energy is proportional to the
interfacial area $A$, plus terms for bulk pressure-volume work. The  pressure difference between the outside and inside  of the membrane is $\Delta P$. 

The Yang-Laplace free energy

\begin{equation}
F=\oint \gamma (\overrightarrow{n})dA+\Delta P\int dV.  \label{7}
\end{equation}

For a system with a fixed volume and external conditions, equilibrium
corresponds to a minimum in $F$. Then

\begin{equation}
\delta F=0,\qquad H=\frac{\Delta P}{2\gamma }=-\frac{1}{R}.  \label{8}
\end{equation}%
This is the Young-Laplace equation. The difference in sign from the force
balance derivation is purely a convention on the sign of $H$; here, $H$ is
often taken with the convention that the mean curvature of a sphere is
negative if the normal points outwards. The physical meaning is identical:
the pressure is higher on the concave side.

Alexandrov rigorously proved that if a smooth, compact (closed and bounded),
and embedded surface in three-dimensional Euclidean space $R^{3}$ has
constant mean curvature ($H=const$), then it must be a sphere \cite{8}.

The quest to understand and predict the equilibrium shapes of
interfaces---from soap films and liquid droplets to biological membranes and
crystalline solids---constitutes a cornerstone of materials science and soft
matter physics. The modern membrane shape equation is the culmination of
centuries of thought, elegantly synthesizing principles from geometry,
thermodynamics, and elasticity. This section traces the critical historical
contributions of N. Stensen, J. Plateau, T. Young, P.S. Laplace, and G.
Wulff.

\section{Shape of red blood cell and elastic theory of membranes in liquid
crystal phases: Helfrich Model for bilayer vesicles}

Biological membranes are essentially two-dimensional fluid sheets composed
of lipid molecules and proteins. Their mechanical properties -- bending
rigidity, spontaneous curvature, and surface tension -- govern a wide range
of cellular processes, from vesicle formation to cell adhesion and division.
In 1973, Wolfgang Helfrich proposed a phenomenological curvature-elasticity
free energy for a fluid membrane, which has become the standard theoretical
tool for describing membrane shapes \cite{1B}.

\subsection{Shape problem of red blood cell}

The red blood cell (RBC) is the unique cell without nucleus in human body.
its shape depends on the cell membranes and environment in physiology. The
red blood cell, with its elegant biconcave disc shape, is a marvel of
biological engineering. Why the RBCs in human body are always in a
rotationally symmetric and biconcave, neither convex nor spherical? For
centuries, this simple, anucleate cell has captivated scientists, driving
research that spans physics, chemistry, and medicine. The journey to
understand the RBC encapsulates a microcosm of scientific progress, from
early descriptive studies to quantitative biophysical modeling and,
recently, to the discovery of its surprising non-canonical functions.

E. Ponder was a pivotal figure in early RBC science. His work systematically
characterized the reversible transformation of RBCs from the normal
discocyte shape to cremated spheres and smooth spheres \cite{10}. He
explored the role of plasma factors and the metabolic energy (ATP) required
to maintain the biconcave shape. He also found high deformability of RBC for
transportation of oxygen in capillary blood vessels. His most enduring
technical contribution is "Ponder's Rule", a method formulated in 1930 to
accurately determine the true location and geometry of the RBC membrane from
light microscopy images, a significant challenge due to optical diffraction
limits \cite{11}.

Building on phenomenological observations, Y.C. Fung and P. Tong provided a
rigorous quantitative and theoretical framework. They studied the thickness
of membrane varied from region to region to regulate the biconcave shape,
contradictory to observation under electron microscope (EM). They also
proposed a mechanical model to explain the shape transformation in 1968 \cite%
{10}. Their work was instrumental in bridging descriptive biology with
engineering principles.

J.R. Murphy made significant contributions to this area in the 1960s and
70s. He developed a density-based separation technique that allowed
researchers to fractionate RBCs by age. This method, often referred to as
"separation according to Murphy," became instrumental in studying RBC aging.
Using this technique, researchers demonstrated that critical metabolites
like adenosine triphosphate (ATP) exhibit a cell-age-dependent decrease, a
finding critical for understanding the metabolic lifespan and eventual
removal of RBCs from circulation. Murphy's work provided the biochemical and
methodological tools to link cellular aging with physical and functional
changes \cite{12}.

Research also delved into the genetic regulation of RBC production
(erythropoiesis) and function. The work of L. Lopez and colleagues in the
early 2000s exemplified this molecular approach. They investigated the role
of the Ikaros transcription factor in hematopoiesis. Using Ikaros null mice,
they demonstrated that this factor is crucial for multiple stages of blood
cell development and for the proper timing of hemoglobin switching from
fetal to adult forms \cite{13}. This highlighted the complex genetic
circuitry that governs the creation of a functional RBC.

The seminal work of Greer \& Baker (1970) provided a foundational physical
principle. They proposed that the biconcave shape could be explained by the
principle of minimum bending energy. They argued that, given a constant
surface area and volume (constraints inherent to the RBC membrane), the
shape that minimizes the curvature elastic energy of the membrane is the
biconcave disc \cite{14}. This was a crucial conceptual leap, framing the
RBC not just as a biological bag but as a physical object obeying mechanical
laws.

This idea was mathematically formalized and profoundly expanded by Wolfgang
Helfrich in the early 1970s. Helfrich introduced his now-famous spontaneous
curvature model for lipid bilayers, treating the RBC membrane as a
two-dimensional liquid crystal. The Helfrich free energy describes the cost
of bending the membrane and includes a parameter for its intrinsic or
"spontaneous" curvature. This model successfully explained not only the
stable biconcave shape but also the various morphological transformations
that RBCs undergo under chemical or physical stress. The "Zhong-Can-Helfrich
equation," developed from this theory, became a cornerstone for modeling RBC
and vesicle shapes. This work elegantly bridged soft matter physics and cell
biology, demonstrating that the RBC's form is a direct physical consequence
of its material properties.

\subsection{Helfrich model for elasticity of lipid bilayers derived by
liquid crystal Curvature Elastic Energy}

F.C. Frank laid the rigorous continuum mechanical foundation for the theory
of nematic liquid crystals in 1958. He formally established the description
of the elastic forces that resist distortions in the molecular alignment of
a nematic, a state of matter with long-range orientational order but no
positional order.

The most commonly used liquid crystals (LCs) are uniaxial anisotropic fluids
formed by rod-like molecules oriented with their longest molecular axes
along an average direction called the director, which is a unit vector $%
\overrightarrow{d}$.

The Frank free energy of curvature elasticity in liquid crystal (LC)

\begin{equation}
F=\oint g_{LC}dA,  \label{9}
\end{equation}%
where free energy density \cite{15}%
\begin{eqnarray}
g_{LC} &=&\frac{1}{2}[k_{11}(\nabla \cdot \overrightarrow{d}%
-S_{0})^{2}+
k_{22}(\overrightarrow{d}\cdot \nabla \times \overrightarrow{d}%
-k_{2}/k_{22})^{2}+k_{33}(\overrightarrow{d}\cdot \nabla \overrightarrow{d}%
)^{2}\\
&&-k_{12}(\nabla \cdot \overrightarrow{d})(\overrightarrow{d}\cdot \nabla
\times \overrightarrow{d})]  \notag -\frac{1}{2}(k_{22}+k_{24})[(\nabla \cdot \overrightarrow{d})^{2}+\\
&&(\nabla
\times \overrightarrow{d})^{2}-\nabla \overrightarrow{d}\colon \nabla 
\overrightarrow{d}].  \label{11}
\end{eqnarray}

Here $k_{11}$, $k_{33}$, and $k_{22}$ are the respective elastic moduli, the 
$S_{0}$ is spontaneous splay.

For achiral (Nematic) LC 
\begin{equation}
k_{2}=0.\text{ }  \label{12}
\end{equation}

For chiral (Cholesteric) LC%
\begin{equation}
k_{2}\neq 0.  \label{13}
\end{equation}

Before Helfrich, various models failed to fully explain the stable biconcave
shape of red blood cells (discocytes). Helfrich's breakthrough was
conceptual: recognizing the membrane as a liquid-crystalline sheet rather
than a simple isotropic fluid or solid shell.

Helfrich's key insight was to treat the cell's lipid bilayer as a
two-dimensional fluid in a liquid crystal (smectic) phase. By doing so, he
could apply the principles of liquid crystal elasticity---pioneered by F.C.
Frank in 1958---to describe the membrane's resistance to bending.

Helfrich proposed that the curvature elastic energy per unit area of a fluid
membrane is given by:

\begin{equation}
F=\oint gdA,  \label{14}
\end{equation}%
where 

\begin{equation}
g=\frac{1}{2}k_{c}(2H+C_{0})^{2}+k_{G}K.  \label{15}
\end{equation}%
Here $H$ is mean curvature (local measure of how much the surface bends), K
is Gaussian Curvature (related to the surface's topology), $k_{c}$ is the
bending rigidity (stiffness against bending), $k_{G}$ is Gaussian curvature
modulus (influences topology changes; hard to measure directly), $C_{0}$ is
spontaneous curvature (most innovative term; represents the membrane's
intrinsic tendency to curve due to asymmetry between its two lipid layers).

Compairing the\ Eq. (\ref{11}) with Eq. (\ref{14}), it is obtained that%
\begin{equation}
k_{c}=k_{11}t  \label{16}
\end{equation}

\begin{equation}
k_{G}=-(k_{22}+k_{24})t  \label{17}
\end{equation}

\begin{equation}
C_{0}=S_{0}/t  \label{18}
\end{equation}

The elastic moduli $k_{11}$ is about $10^{-11}$ N, and the membrane thickness  $t$ is about $5$ nm.
Then the bending module $k_{c}$ is about $10^{-12}$ $erg$.

The osmotic pressure outside the membrane is $P_{o}$, the osmotic pressure inside the membrane is $P_{i}$.  Then the osmotic pressure difference between the outside and inside  of the membrane is

\begin{equation}
\Delta P=P_{o}-P_{i} . \label{6}
\end{equation}
The tensile stress acting on the membrane is defined as $\lambda$ . 

Helfrich free energy is written as \cite{1B}

\begin{equation}
F=\oint gdA+\Delta P\int dV+\lambda \oint dA.  \label{19}
\end{equation}%
Mathematically, $\Delta P$  and $\lambda$ can be considered as Lagrange multiplies.

\subsection{Zhong-Can-Helfrich membrane shape equation}

The shape that minimizes this total energy is the predicted stable form.
Minimizing this functional leads to a complex shape equation, solutions to
which successfully describe the biconcave discoid and other shapes.

In 1987, Zhong-Can Ou-Yang and Helfrich minimized this total energy and
 obtained the Zhong-Can-Helfrich equation \cite{1C}

\begin{equation}
\Delta P-2\lambda H+k_{c}(2H+C_{0})(2H^{2}-2K-C_{0}H)+2k_{c}\nabla ^{2}H=0,
\label{23}
\end{equation}

Here%
\begin{equation}
\nabla ^{2}=(1/\sqrt{g})\partial _{i}(g^{ij}\sqrt{g}\partial _{j}).
\label{25}
\end{equation}

is the generalized Laplace formula \cite{18}-\cite{19}.

The membrane shape equation (\ref{23}) is a Laplace Equation. Alexandrov
rigorously proved that if a smooth, compact, and embedded surface in
three-dimensional Euclidean space $R^{3}$ has constant mean curvature ($%
H=const$), then it must be a sphere. But Helfrich Variation have $3$ cases
for sphere solutions governed by

\begin{equation}
\Delta P\cdot r_{0}^{3}+2\lambda r_{0}^{2}-kC_{0}r_{0}(2-C_{0}r_{0})=0,
\label{26}
\end{equation}

Large molecules such as proteins cannot directly traverse the lipid bilayer
of the cell membrane. Their transport primarily relies on vesicular
trafficking. This process depends on specific membrane proteins (e.g., SNARE
proteins) to recognize, dock, and fuse vesicles. The three spherical
solutions of the membrane shape equation (\ref{26}) can be applied to model
and compute the cellular processes of exocytosis and endocytosis.
Endocytosis and exocytosis are the universal pathways for the exchange of
macromolecules between the cell and its external environment. Autophagy is a
highly specific, internally-directed special form of "endocytosis" regulated
by a complex genetic network. Research on membrane transport and autophagy
has given rise to biotechnological frontiers such as transmembrane protein
delivery, which aims to utilize or mimic these mechanisms for drug delivery.
Yoshinori Ohsumi was awarded the 2016 Nobel Prize in Physiology or Medicine
for elucidating the molecular mechanisms of autophagy \cite{20}. If the Eq. (%
\ref{26}) has two positive roots, then these two roots can be used to
characterize the radii of the two daughter cells in cell division, as well
as the phagophore and the vesicle in exocytosis. If the Eq. (\ref{26}) has
one positive root and one negative root, then these two roors can be used to
characterize the phagocytic vesicle and the endocytic vesicle in
endocytosis. The Helfrich model explains the shape and plasticity of
membranes from a physics perspective, which serves as the structural
prerequisite for all membrane transport processes, including autophagosome
formation.

According to tradition, the coordinate $(X,Y,Z)$ of a point on the membrane
surface is choosed. If the surface is axisymmetric, the symmetric axis is
the $Z$-axis. The radial distance from the surface point to the $Z$-axis is $%
\rho $. The angle between the projection of the point onto the base $X-Y$
plane and positive $X$-axis is defined as the azimuthal angle $\theta $. The
angle between the tangent line of the contour at that point and the $\rho $%
-axis is $\psi $. Then, $X=\rho \cos (\theta )$, $Y=\rho \sin (\theta )$,
and $dZ=tan(\psi )d\rho $.

A major breakthrough by Zhong-Can Ou-Yang and J.G. Hu was to specialize this
general shape equation (\ref{23}) to the case of axisymmetric shapes, shapes
possessing rotational symmetry around an axis, such as spheres, tubes, or
biconcave discs. Shape equation of rotationally symmetric vesicles is \cite{21}%

\begin{eqnarray}
\cos ^{3}\psi \frac{d^{3}\psi }{d\rho ^{3}} &-&4\sin \psi \cos ^{3}\psi 
\frac{d^{2}\psi }{d\rho ^{2}}\frac{d\psi }{d\rho }+\cos \psi (\sin ^{2}\psi -%
\frac{1}{2}\cos ^{2}\psi )(\frac{d\psi }{d\rho })^{2}-\frac{7\sin \psi \cos
^{2}\psi }{2\rho }(\frac{d\psi }{d\rho })^{2}  \notag \\
&&+\frac{2\cos ^{2}\psi }{\rho }\frac{d^{2}\psi }{d\rho ^{2}}-[\frac{%
C_{0}^{2}}{2}+\frac{\lambda }{k_{c}}+\frac{2C_{0}\sin \psi }{\rho }-\frac{\sin ^{2}\psi -2\cos
^{2}\psi }{2\rho ^{2}}]\cos \psi \frac{d\psi }{d\rho }  \notag \\
&&-[-\frac{\Delta P}{%
k_{c}}+\frac{\lambda \sin \psi }{k_{c}\rho }+\frac{C_{0}^{2}\sin \psi }{2\rho }-\frac{\sin \psi }{2\rho ^{3}%
}-\frac{\sin \psi \cos ^{2}\psi }{2\rho ^{3}}]=0.
\label{28}
\end{eqnarray}

Previous shape equation of rotationally symmetric vesicles the special case
of the zero constant of the first integral of the shape equation, a 1D plane
curve variation may induce special 2D curved surface variation \cite{22}%
\begin{eqnarray}
\cos ^{2}\psi \frac{d^{2}\psi }{d\rho ^{2}}-\frac{1}{2}\cos \psi \sin \psi (%
\frac{d\psi }{d\rho })^{2} +\frac{\cos ^{2}\psi }{\rho }\frac{d\psi }{d\rho }-%
\frac{\sin 2\psi }{2\rho ^{2}} \notag \\
-\frac{\overline{\Delta P}\rho }{2}\cos \psi -%
\frac{\lambda \sin \psi }{k_{c} \cos \psi }- 
\frac{\sin \psi }{2\cos\psi }(\frac{\sin \psi }{\rho }+C_{0}^{2})=0,  \label{29}
\end{eqnarray}%
where $\overline{\lambda }=\lambda /k_{c}+C_{0}^{2}/2$, $\overline{\Delta P}%
=\Delta P/k_{c}$.

The shape of the red blood cell has been computed using the above elasticity
model (Deuling \& Helfrich, 1976 \cite{22}; Jenkins, 1977 \cite{23};
Peterson, 1985 \cite{24}; Svetina \& \v{Z}ek\v{s}, 1989 \cite{25}; Miao et
al., 1991 \cite{26}).

Another special variational method is the 1D plane curve variation

\begin{equation}
\delta \int F(\psi (s),d\psi /ds,d^{2}\psi /ds^{2})ds=0.  \label{31}
\end{equation}

An special membrane shape equation was derived by this variational method
(Berndl et al, 1990 \cite{27}; Seifert 1991 \cite{28}; Seifert et al., 1991 
\cite{29}, Julicher, 1994 \cite{30}).%
\begin{eqnarray}
\cos ^{3}\psi \frac{d^{3}\psi }{d\rho ^{3}} &-&(3\sin \psi \cos ^{2}\psi +%
\frac{\cos ^{2}\psi }{\sin \psi })\frac{d^{2}\psi }{d\rho ^{2}}\frac{d\psi }{%
d\rho }+\cos \psi \sin ^{2}\psi (\frac{d\psi }{d\rho })^{3}  \notag \\
&&-\frac{(2+5\sin ^{2}\psi )\cos ^{2}\psi }{2\rho \sin \psi }(\frac{d\psi }{%
d\rho })^{2}+\frac{2\cos ^{3}\psi }{\rho }\frac{d^{2}\psi }{d\rho ^{2}}+
\notag \\&& [%
\frac{C_{0}\sin \psi }{\rho }+\frac{\sin ^{2}\psi }{\rho ^{2}}+\frac{\Delta
P\rho }{2k\cos \psi }]\cos \psi \frac{d\psi }{d\rho }  \notag \\
&&-[\frac{\Delta P}{k}+\frac{\lambda \sin \psi }{k\rho }+\frac{C_{0}^{2}\sin
\psi }{2\rho }-\frac{\sin \psi (1+\cos ^{2}\psi )}{2\rho ^{3}}]=0.  \label{32}
\end{eqnarray}

Many authors have accepted the biomembrane shape equation (\ref{28}) in 2002 
\cite{31}.

The first integral of shape equation (\ref{28}) is found by W. M. Zheng and
J. Liu \cite{32}:%
\begin{equation}
\frac{\Psi ^{3}-\Psi (\rho \Psi ^{\prime })^{2}}{2\rho }-\rho (1-\Psi ^{2})[%
\frac{(\rho \Psi )^{\prime }}{\rho }]^{\prime }-C_{0}\Psi ^{2}+\overline{%
\lambda }\rho \Psi +\overline{\lambda }\rho \Psi +\frac{\overline{\Delta P}%
\rho ^{2}}{2}=\eta _{0},  \label{33}
\end{equation}%
where $\Psi =\sin \psi $.

\subsection{Solution of Shape equation of vesicles with rotationally
symmetric.}

For over a century, the unique biconcave disc shape of the human red blood
cell (RBC or discocyte) has been a subject of fascination and a major puzzle
in biophysics. This shape is crucial for the cell's function, optimizing gas
exchange and deformability. Before the 1990s, while the Helfrich curvature
elasticity model successfully explained the shapes of simple lipid vesicles,
a first-principles, purely mechanical derivation of the specific RBC contour
from a single, closed-form mathematical equation remained elusive.

The analysis is built upon the Helfrich theory of fluid membrane. The lipid
bilayer is treated as a two-dimensional liquid crystal sheet, with its
equilibrium shape determined by minimizing the bending elastic energy.
Naito, Okuda, and Ou-Yang's primary contribution \cite{1E} was to
demonstrate that under the condition of zero osmotic pressure difference, the
axisymmetric shape equation admits a specific analytic solution that
perfectly describes the classic biconcave disc profile. This solution is not
a numerical approximation but a closed-form mathematical function:

\begin{equation}
\sin \psi =C_{0}\rho \ln (\rho /\rho _{B}),  \label{34}
\end{equation}%
where $C_{0}<0$ for bioconcave shape of RBC.

The radius $R_{0}$ of RBC is computed as \cite{34}

\begin{equation}
R_{0}=\sqrt{A/4\pi }=3.25\text{ }\mu m,  \label{35}
\end{equation}%
and

\begin{equation}
C_{0}R_{0}=(\frac{\sqrt{5}-1}{2})^{-1}\text{.}  \label{36}
\end{equation}

These theoretical results are in excellent agreement with the experimental
measurements \cite{35}.

Phenomenologically, we may asuume that the complex agents of the asymmetry
of the membrane produce an electric field $\overrightarrow{E}$:%
\begin{equation}
\overrightarrow{E}=-\Delta \psi \overrightarrow{n}/d,  \label{37}
\end{equation}%
where electric potential of membrane

\begin{equation}
\Delta \psi =\psi _{0}-\psi _{i}<0.  \label{38}
\end{equation}

Analogous to piezoelectricity in solids, Meyer introduces the concept of
curvature electricity \cite{36} in the field of LC. The polarization induced
by bending is

\begin{equation}
P=e_{11}\overrightarrow{n}\nabla \cdot \overrightarrow{n}=e_{11}%
\overrightarrow{n}(-2H),  \label{39}
\end{equation}%
where $e_{11}$ is the flexoelectric constant.

This effect gives rise to an additional membrane energy:

\begin{equation}
\Delta F=-\oint dA\int\nolimits_{0}^{d}\overrightarrow{P}\cdot 
\overrightarrow{E}dz.  \label{40}
\end{equation}

Incorporation of Eq. (\ref{15}) and Eq. (\ref{40}) gives

\begin{equation}
C_{0}=e_{11}\Delta \psi /k_{c}.  \label{41}
\end{equation}

The flexoelectric constant and the bending constant can be estimated as \cite{37}%
, \cite{38}

\begin{equation}
e_{11}\simeq 10^{-4}dyne^{1/2},\qquad k_{c}\simeq 10^{-12}erg.  \label{42}
\end{equation}

Then the red blood cell potential

\begin{equation}
\Delta \psi \simeq -15.0mv.  \label{43}
\end{equation}

This theoretical cell potential fits very well with experimental result $%
\Delta \psi =-14.0mv$ which is measured by Lassen et al. \cite{39}.

The 1991's textbook "Molecular and Cell Biophysics" \cite{40} written by
R.J. Nossal \& H. Lecar regards W. Helfrich liquid crystal curvature
elasticity model of membranes as the interpretation of RBC shape. The work 
\cite{1E}, \cite{34} by Ou-Yang et al. resolved a fundamental question in
cell morphology. It provided the long-sought direct link between the
abstract Helfrich theory and the concrete, well-known morphology of the RBC,
closing a key loop in theoretical biophysics. The analytic solution served
as a precise baseline for subsequent investigations into RBC mechanics,
stability, and transformations. It allowed researchers to analytically
calculate stresses and to perturb the solution to study shape transitions
under changes in volume or other parameters.

The equilibrium shapes of closed vesicles were primarily understood to be
spheres, prolates, oblates, or discocytes (like red blood cells). Within the
framework of Helfrich's theory for fluid membranes, Ou-Yang et al. first
predicted the solution of a toroidal vesicle shape in 1990 \cite{41}:

\begin{equation}
\sin \psi =\rho /r-\sqrt{2},  \label{44}
\end{equation}%
which under constraint condition

\begin{equation}
C_{0}r<(-\frac{\sqrt{3}}{2}-\frac{\sqrt{2}}{4})\pi ^{1/2}(\sqrt{2})^{-1/2}.
\label{45}
\end{equation}

This torus solution of membrane is a direct and brilliant application of the
Zhong-Can-Helfrich equation on the axisymmetric shape equation. It showcases
the power of the analytical approach to solve nonlinear problems in membrane
mechanics. Furthermore, the stability analysis of the torus is intimately
connected to the study of spontaneous curvature $C_{0}$. The paper discusses
how the stability condition interacts with $C_{0}$, linking this exotic
shape to the same fundamental material property that governs red blood cell
transformations. This work is confirmed by experiments \cite{42}-\cite{44}.

\subsection{Polygon deformation instability of spherical vesicle and myelin form}

In 1987, Z.C. Ou-Yang and W. Helfrich have generalized the work done by W.
Helfrich in 1973. They have treated the sphere as the "ground state" of a
closed vesicle under a reference pressure. He then introduced infinitesimal
shape perturbations, mathematically expressed as a series of spherical
harmonics $Y_{lm}(\theta ,\phi )$. Each harmonic $Y_{lm}(\theta ,\phi )$
represents a distinct deformation mode: The index $l$ (degree) determines
the wavelength or "wiggliness" of the deformation around the sphere.\ The
index $m$ (order) describes the azimuthal orientation of the deformation. By
substituting the perturbed shape into the Helfrich free energy functional
and keeping terms to second order, the problem reduces to an eigenvalue
problem. The eigenvalues determine the critical pressure at which the
spherical state becomes unstable against a perturbation of a given $l$ mode.
A negative eigenvalue indicates instability.

The critical pressure \cite{1C}

\begin{equation}
\Delta P\equiv \frac{2k_{c}}{r_{0}^{3}}[l(l+1)-C_{0}r_{0}]  \label{46}
\end{equation}

When $l=2$, the critical pressure is calculated by W. Helfrich \cite{1B}

\begin{equation}
\Delta P\equiv \frac{2k_{c}}{r_{0}^{3}}(6-C_{0}r_{0}).  \label{47}
\end{equation}%
it describes an axisymmetric quadrupole deformation, i.e., precisely the
kind of symmetric "pinching" at two poles that can transform a sphere into
an oblate, prolate, or biconcave shape. This mode is identified as the
primary instability leading to the discocyte.

When $l=3$ or $l=4$ modes could become unstable first. These correspond to
more complex, non-axisymmetric deformations, which are relevant for
understanding pathological or chemically induced red cell shapes (e.g.,
echinocytes with multiple spicules).

Their work has marked a critical transition in membrane biophysics: moving
from the search for static equilibrium shapes to the analysis of their
dynamic stability and the pathways of shape transformation. Their work
provides the formal theoretical machinery to understand how and why a
spherical red blood cell (or any lipid vesicle) can transform into a
biconcave discocyte or other complex forms \cite{46}.

The configuration of red blood cells is usually similar to that of a
biconcave disk in human beings and sphere in breastfed animals. But when red
cells are profoundly damaged or become necrotic \cite{47}, the interaction
of phospholipids with the aqueous protein solution gives rise to structures
called myelin figures \cite{48}. They are of two types: external and
internal structures.

Recently, the phenomenon of budding, i.e., the expulsion of a smaller
vesicle out of a larger one, has attracted a lot of interest \cite{49}. Many
response experiments of single-component giant unilamellar vesicles (GUVs)
subjected to different external osmotic stresses have been made. It has been
found that giant vesicles can transform into different multispheres; the
small spheres can be external or internal. This process is similar to the
myelin formation of red blood cells. In fact, these shape transformations
can be unified, described, and computed using the membrane theory of Ou-Yang
Zhong-Can and Helfrich \cite{1C}.

Zhou J.J. et al \cite{50} have used the perturbation of sin$\psi $ and
software Surface Evolver to compute Myelin Form happened in death of RCB.
They found the conical function $P_{-\frac{1}{2}+iq}$ can be used to
describe the deformation of red blood cell where

\begin{equation}
l=-\frac{1}{2}+iq.  \label{48}
\end{equation}

In 2025, T. Xu and Z.C. Ou-Yang have constructed a multiple solutions theory 
\cite{51} based on a membrane shape equation. These spherical solutions of
Zhong-Can--Helfrich shape equation has spherical solutions in a line. These
spheres have an identical radius $r_{s}$ but different center positions, can
be described by the same equation: $\phi -\rho /r_{s}=0$. Therefore, there
can be multiple solutions for the sphere equilibrium shape equation, and
these need to satisfy a quadratic equation. The quadratic equation has a
maximum of two nonzero roots for Eq. (\ref{26}). Then the multiple solutions
can be written as%
\begin{equation}
(\phi -\rho /r_{s1})^{N_{1}}(\phi -\rho /r_{s2})^{N_{2}}=0,  \label{48B}
\end{equation}%
where $N_{1}$ is the number of spheres with radius $r_{s1}$ and $N_{2}$ the
number of spheres with radius $r_{s2}$.

Radii $r_{s1}$ and $r_{s2}$ satisfy the relation%
\begin{equation}
\frac{1}{r_{s1}}+\frac{1}{r_{s2}}=\frac{\overline{\lambda }}{c_{0}}
\label{48C}
\end{equation}

The multiple sphere solutions should be in a line to undergo rotational
symmetry. The quadratic equation is used to compute the sphere radius,
together with a membrane surface constraint condition, to obtain the number
of small spheres. Matching with the energy constraint condition to determine
the stability of the full solutions. The method is then extended into the
Myelin formation of red blood cells. Their numerical calculations show
excellent agreement with the experimental results and enable the
comprehensive investigation of cell fission and fusion phenomena.
Additionally, the existence of the bifurcation phenomenon in membrane growth
is predicated and proposed a control strategy. For more theoretical analysis
and experimental results, one can refer to R. Lipowsky's works in References 
\cite{52}, \cite{53}.

Helfrich membrane theory, under a given set of physical
conditions---specifically constant osmotic pressure and membrane
tension---provides a framework capable of describing a range of red blood
cell morphologies, notably the classic biconcave disc and myelin sheath-like
configuration.

\subsection{Anharmonic magnetic deformation of spherical vesicle:
Field-induced tension and swelling effects}

M. Iwamoto, Z-C Ou-Yang have investigated the theoretical deformation of a
spherical lipid vesicle when placed in a uniform magnetic field \cite{54} in
2013. The membrane is treated as a two-dimensional magnetic medium with
diamagnetic anisotropy, meaning its response to the magnetic field depends
on the orientation of its molecules.

The total free energy of the vesicle is postulated as the sum of the bending
energy $F_{b}$ and a magnetostriction energy $F_{H}$:

\begin{equation}
F=F_{b}+F_{H}+k_{G}\oint KdA+\Delta P\int dV+\lambda \oint dA,  \label{49}
\end{equation}%
where

\begin{equation}
F_{b}=\frac{1}{2}k_{c}\oint (C_{1}+C_{2}-C_{0})^{2}dA,  \label{50}
\end{equation}%
and

\begin{equation}
F_{H}=-\frac{1}{2}\Delta \chi t\oint (\overrightarrow{\mathbf{H}}\cdot 
\overrightarrow{n})^{2}dA.  \label{50B}
\end{equation}

The variation of magnetostriction energy

\begin{equation}
\delta F_{H}=-\Delta \chi t\oint \overrightarrow{[\mathbf{H}}(%
\overrightarrow{H}\cdot \overrightarrow{n})^{2}+\nabla \cdot (%
\overrightarrow{\mathbf{H}}\cdot (\overrightarrow{\mathbf{H}}\cdot 
\overrightarrow{n}))]\psi dA.  \label{51}
\end{equation}

Then the membrane shape equation with magnetic field is

\begin{equation}
\begin{split}
\Delta P-2\lambda H+k_{c}(2H+C_{0})(2H^{2}-2K-C_{0}H)+2k_{c}\nabla
^{2}H \\
=\Delta \chi t\oint \overrightarrow{[\mathbf{H}}(\overrightarrow{%
\mathbf{H}}\cdot \overrightarrow{n})^{2}+\nabla \cdot (\overrightarrow{%
\mathbf{H}}\cdot (\overrightarrow{\mathbf{H}}\cdot \overrightarrow{n}))]\psi
dA  \label{52}
\end{split}
\end{equation}

Finally the field-induced normalized birefringence is obtained and the
change of R:

\begin{equation}
\frac{\Delta n}{\Delta n_{\max }}\approx \frac{R(\theta =90)-R(\theta =0)}{R}%
=\frac{\overline{\beta }H_{0}^{2}}{1+\eta \overline{H}_{0}^{2}}.  \label{53}
\end{equation}

This work provides a first-principles explanation for how static magnetic
fields can directly and non-invasively alter both the shape and volume of
cell-like membranous objects. The predicted swelling effect is a
non-trivial, testable phenomenon relevant for understanding bio-magnetic
effects and designing magnetic control strategies in biophysics. One can
obtain membrane shape equuation with electric field by substituting $\Delta
\chi H$ for $\Delta \varepsilon E$ by the same method.

The 1998 study by Saitoh et al. (PNAS) \cite{55} was a groundbreaking
experimental discovery. Using high-resolution microscopy and biochemical
experiments, they demonstrated that the membrane protein Talin could induce
stabilized openings in liposomes, thereby challenging the long-held dogma
that lipid bilayers cannot maintain free edges in water.

Z.C. Tu and Z.C. Ou-Yang have derived the governing shape equations directly
from differential geometry principles \cite{56}-\cite{58}. The osmotic
pressure $\Delta P$ equals zero, then the shape equations for open vesicles
is

\begin{equation}
k_{c}(2H+C_{0})(2H^{2}-2K-C_{0}H)-2\lambda H+2k_{c}\nabla ^{2}H=0.
\label{54}
\end{equation}

The points on the boudary should satisify the constraint conditions

\begin{equation}
\lbrack (2H+C_{0})+\widetilde{k_{G}}k_{n}]|_{C}=0,  \label{55}
\end{equation}

\begin{equation}
-[2\overrightarrow{b}\cdot \nabla H+\widetilde{\gamma }k_{n}+\widetilde{k_{G}%
}\frac{d\tau _{g}}{ds}]|_{C}=0,  \label{55B}
\end{equation}

and

\begin{equation}
\lbrack \frac{1}{2}(2H+C_{0})^{2}+\widetilde{k_{G}}K+\widetilde{\lambda }+%
\widetilde{\gamma }k_{\theta }]|_{C}=0.  \label{56}
\end{equation}%
Here $\widetilde{\lambda }\equiv \lambda /k_{c}$, $\widetilde{k_{G}}\equiv
k_{G}/k_{c}$, $\widetilde{\gamma }\equiv \gamma /k_{c}$ are the reduced
surface tension, reduced bending modulus, and reduced line tension. $k_{n}$, 
$k_{g}$,and $\tau _{g}$ are the normal curvature, geodesic curvature, and
geodesic torsion of the boundary curve, respectively. $\frac{d\tau _{g}}{ds}$
represents the derivative with respect to the arc length of the open
membrane boudary \cite{59}. These results naturally recover the
Zhong-Can-Helfrich equation while being more powerful for analyzing complex topologies
and constraints.

The work by Z.C. Ou-Yang and Z.C. Tu is a theoretical model construction. It
gave a phenomenogical explanation to the experiment result of Saitoh et al. 
\cite{55}, and established a phase-field free energy functional based on
thermodynamics and statistical physics, creating a unified molecular theory
to model membrane stability, thickness variation, and microdomain formation.

Now, let us discuss the dynamics of vesicles and micro-emulsion droplets in
liquid such as blood. For incompressible fluids in the presence of
conservative body force fields, the continuity equations become

\begin{equation}
\nabla \cdot \overrightarrow{V}_{out,in}=0,  \label{57}
\end{equation}%
where $\overrightarrow{\mathbf{V}}_{out,in}$ is the local mass average fluid
velocity, and measured relative to an inertial frame.

Navie-Stokes equation

\begin{equation}
\rho _{b,out,in}(\frac{\partial }{\partial t}\overrightarrow{V}_{out,in}+%
\overrightarrow{V}_{out,in}\cdot \nabla \overrightarrow{V}_{out,in})=-\nabla
P_{out,in}+\mu _{out,in}\nabla ^{2}\overrightarrow{V}_{out,in},  \label{58}
\end{equation}%
where $\rho _{b}$ is the local density of the fluid, such as blood in human
being, the fluid's viscosity is $\mu $.

Now, let us consider the low Reynolds number flow of blood in the vessels. A
uniform velocity $U$ of red blood cells relative to blood in the capillaries
is assumed, and the blood's viscosity is $\mu $. The blood can be
approximated as Newtonian fluid. The stresses on the fluid element per unit
volume of incompressible blood are denoted as%
\begin{equation}
\mathbf{\Pi }=-P_{b}\mathbf{I}+\mu \lbrack \nabla \overrightarrow{V}+(\nabla 
\overrightarrow{V})^{T}],  \label{59}
\end{equation}%
where $\overrightarrow{\mathbf{V}}$ is the velocity of blood and $P_{b}$ is
the blood pressure caused by the relative motion of cells. The sphere-like
red-blood cells have obtained external radial stress $\mathbf{\sigma }_{bn}$
along the radial direction $\mathbf{n}$ 
\begin{equation}
\mathbf{\sigma }_{bn}=\mathbf{\Pi \colon }\overrightarrow{n}\mathbf{.}
\label{60}
\end{equation}

The radial force balance on a cell membrane requires that the total radial
force should be zero. The radial stress difference $\Delta \sigma $ is
defined as 
\begin{equation}
\Delta \sigma _{bn}=\sigma _{bn,out}-\sigma _{bn,in},  \label{60B}
\end{equation}%
where $\sigma _{bn,out}$ and $\sigma _{bn,out}$ are extracellular and
intracellular radial stress $\mathbf{\sigma }_{bn}$ of the membrane.

Then the general membrane shape equation of cell in blood is

\begin{equation}
\Delta P-2\lambda H+k(2H+C_{0})(2H^{2}-2K-C_{0}H)+2k\nabla ^{2}H=\Delta
\sigma _{bn},  \label{61}
\end{equation}

Sphere-like red blood cells are acted upon by external blood pressure \cite%
{2}, \cite{60} 
\begin{equation}
P_{b}=1.5\mu U\cos (\theta )/r.  \label{62}
\end{equation}%
Here, $r$ is the radius of the cell, and $\theta $ is the angle from the $Z$%
-axis. The external blood pressure $P_{b}$ is about 10$^{-6}\cos (\theta )/r$
in capillaries. The shape equation becomes \cite{51}%
\begin{equation}
\overline{\Delta P}-\frac{2C_{0}}{r^{2}}+\frac{2\overline{\lambda }}{r}=%
\overline{P_{b}}\text{,}  \label{62B}
\end{equation}%
where $\overline{\Delta P}=\Delta P/k_{c}$, $\overline{P_{b}}=P_{b}/k_{c}$.
Then $\overline{P_{b}}=\frac{1.5\mu U}{k_{c}r}\cos (\theta )$. Consequently,
spherical cells adapt by either deforming to resist the excess pressure by
regulating their surface tension. The shape equation becomes%
\begin{equation}
\overline{\Delta P}-\frac{2C_{0}}{r^{2}}+\frac{2\overline{\lambda _{2}}}{r}=0%
\text{,}  \label{62C}
\end{equation}%
where $\overline{\lambda _{2}}=\lambda /k_{c}+c_{0}^{2}/2-1.5\mu U\cos
(\theta )$. This situation is analogous to perturbations described by
spherical harmonics $Y_{l,m}$, where $l=1$, $m=0$. In steady state, if the
velocity $U$ of red blood cells relative to blood becomes zero, the external
blood pressure $P_{b}$ turns to zero.

\subsection{Theory of helical structure of tilted chiral membranes}

Many biological membranes contain chiral lipid molecules (e.g.,
phospholipids with asymmetric tails) that can exhibit a collective tilt of
molecular chains relative to the membrane normal. Z.C. Ou-Yang and J.X. Liu
have developed theory of helical structures \cite{61}, \cite{62} in tilted
chiral membranes (TCM), which represents a critical expansion into the realm
of molecular chirality and its macroscopic geometric consequences.

The TCM theory introduces a director field $\overrightarrow{d}$ to represent
the average tilt direction of lipid molecules within the tangent plane of
the membrane surface. Chirality is accounted for by a Lifshitz invariant in
the free energy, which favors a continuous rotation (gradient) of the tilt
direction. The total free energy density

\begin{equation}
g_{LC}=\frac{1}{2}k_{11}(\nabla \cdot \overrightarrow{d})^{2}+\frac{k_{22}}{2%
}(\overrightarrow{d}\cdot \nabla \times \overrightarrow{d}-k_{2}/k_{22})^{2}+%
\frac{k_{33}}{2}(\overrightarrow{d}\times \nabla \times \overrightarrow{d}%
)^{2}  \label{63}
\end{equation}

The first approximation of Frank free energy density 
\begin{equation}
g_{LC}=-k_{2}\overrightarrow{d}\cdot \nabla \times \overrightarrow{d}
\label{64}
\end{equation}%
which characterizws the chirality of cholesteric liquid crystal. Here only
this term is used for the strong chirality.

Then the free energy of TCM

\begin{equation}
F=-k_{2}t\cos \theta _{0}\oint \overrightarrow{d}\cdot d\overrightarrow{l}%
-2k_{2}\sin ^{2}\theta _{0}\cos \theta _{0}\int \tau _{g}dA,  \label{65}
\end{equation}%
where geodesic torsion $\tau _{g}$ may be written as 
\begin{equation}
\tau _{g}=(C_{1}-C_{2})\sin \varphi \cos \varphi .  \label{67}
\end{equation}

Here $C_{1}$ and $C_{2}$ are the two principal curvatures and $\theta $ is
angle between the director and the helix surface normal. $\varphi $ is the
angle from one princpal direction to the local tilt direction. The angular
dependencee is just what Helfrich and Prost found in Ref. \cite{63}

Minimizing the above free energy of TCM leads to the local azimuthal angle
of the director must be

\begin{equation}
\varphi =45^{o}  \label{68}
\end{equation}

J. M. Schnur indicated \cite{64}-\cite{66} the result in agreement with
observation but different from those theories proposed by de Gennes \cite{67}%
, Lubensky-Prost \cite{68}.

The micellar model biles composed of bile salt sodium taurocholate,
lecithin, and cholesterol in a molar ratio of $97.5:0.8:1.7$ were initially
prepared and contained both micelles and vesicles. Within $2-4$ hours of
dilution, filamentous structures were observed. A few days later, the
filaments were bent to form high-pitch helices [$\simeq 54^{o}$, Fig. 2(a)
in \cite{69}]. These helices grew laterally while maintaining the pitch
angle to form tubules Within a few weaks, high-pitch helices and tubules
disappeared, while new helices with low-pitch angle ($\simeq 11^{o}$)
appeared and grow to new tubles. This experimental result has confused many
scientists. By taking complete Franck free energy, considering two TCM
helices: ($1$) helical ribbon with parallel packing of molecules; ($2$)
helical ribbon with antiparallel packing of molecules, Komura and Ou-Yang
show \cite{70} the helical angle of ($1$) $<45^{o}$ and that of ($2$) to be%
\begin{equation}
\phi _{0}=arc\tan [\{\frac{8}{3}\cos (\frac{1}{3}\arccos \frac{5}{32}+\frac{1%
}{3})\}^{1/4}]=52.1^{o}.  \label{69}
\end{equation}

The theoretical results are in excellent agreement with the experimental
findings \cite{69}

\begin{equation}
\phi _{0}=53.7\pm 0.8^{o},\qquad \phi _{pal}=11.3^{o}.  \label{70}
\end{equation}

Komura and Ou-Yang's workr \cite{70} is a cornerstone in the physics of
chiral membranes. By elegantly combining Helfrich's membrane mechanics with
Frank's cholesteric elasticity, it delivers a powerful and predictive
theoretical model. The detailed derivation shows how simple geometric
constraints and symmetry-breaking chiral interactions lead to complex,
bistable morphological outcomes---a principle that continues to resonate in
the design of programmable soft materials.

\section{Helfrich Model for Multi-layer Vesicles: Focal conic structures in
Smectic A LC and general variation problem of surfaces.}

Helfrich fluid membrane theory is regarded by J.C.C. Nitsche (1993) \cite{71}
as the renewal of the Poisson's elastic shell theory. Nitsche begins by
associating a free energy per unit area with a surface which depends on the
principal curvatures. Under mild regularity assumptions and the condition
that the functional be definite, he demonstrates that free energy must take
a specific form. By imposing symmetry and definiteness, the energy density
is rigorously reduced to the well-known Helfrich form%
\begin{equation}
F=\oint [\psi (H)-\gamma K]dA,  \label{71}
\end{equation}%
where the free energy functional that is quadratic in the principle
curvatures.

Minimazing the above equation lead to%
\begin{equation}
\nabla ^{2}\psi _{H}+2(2H^{2}-K)\psi _{H}-4H\psi =0,  \label{72}
\end{equation}%
where%
\begin{equation}
\psi _{H}=\frac{\partial \psi }{\partial H}.  \label{73}
\end{equation}

Let $H$, $K$ be the mean and Gaussin Curvatures at inner surface and $D$ be
the thickness of multi-layer vesicle. Taking variation to Helfrich free
energy by this method%
\begin{equation}
\delta F=\delta \oint \Phi (D,H,K)=0,  \label{74}
\end{equation}%
then the most general equation of surface variation for multi-layer vesicles
can be obtained.

Minimazing the Helfrich free energy for surface variation%
\begin{equation}
\delta F/\delta A=0,  \label{75}
\end{equation}%
the most general membrane shape equation is obtained \cite{72}%
\begin{equation}
(2H^{2}-K+\frac{1}{2}\nabla ^{2})\Phi _{H}+(2HK+\overline{\nabla }^{2})\Phi
_{k}-2H\Phi =0.  \label{76}
\end{equation}%
Here%
\begin{equation}
\Phi _{H}=\frac{\partial \Phi }{\partial H},\qquad \Phi _{k}=\frac{\partial
\Phi }{\partial K},  \label{77}
\end{equation}

\begin{equation}
\nabla ^{2}=\frac{1}{\sqrt{g}}\partial _{i}(g^{ij}\sqrt{g}\partial
_{j}),\qquad \overline{\nabla }^{2}=\frac{1}{\sqrt{g}}\partial _{i}(KL^{ij}%
\sqrt{g}\partial _{j}).  \label{78}
\end{equation}

The integral equation can be obtained by 
\begin{equation}
\partial F/\partial D=\oint \partial \Phi /\partial DdA=0.  \label{79}
\end{equation}%
One can find the detailed computation of the integral equation in Ref. \cite%
{72}-\cite{73} which clarify the relationship between the global shape of
the SmA domain and the intricate, nested pattern of its internal layers.
This work establishes a direct, predictable connection between material
parameters (elastic constants) and the resulting domain geometry.

When cooling from the isotropic phase to the smectic \cite{74} A phase, the
system often nucleates by forming a focal conic texture. Within this
texture, the smectic layers are commonly arranged in the form of Dupin
cyclides. This configuration tends toward the minimal energy state by
reconciling the phase transition driving force with the curvature elastic
energy under the condition of layer incompressibility. The relieved energy
of the difference in Gibbs free energy of I-SmA transition must be balanced
by the curvature elastic energy of SmA layers.

W. Bragg \cite{75} raised an important question: Why the cyclides are
preferred to other geometrical structures under the preservation of the
interlayer spacing? H. Naito, M. Okuda, Z.C. Ou-Yang give an answer: The
relieved energy of the difference in Gibbs free energy of I-SmA transition
must be balanced by the curvature elastic energy of SmA layers.

The Sm-A nucleus grown from an I phase can be described as a layer whose
inner and outer surfaces are parallel surfaces. The thickness of the layer
is d. The net difference in the energy between the Sm-A and the I phase is
the sum of three terms:

The first term is the curvature elastic energy change of the Sm-A nucleus:

\begin{equation}
\delta F_{C}=k_{11}d/2\oint (2H)^{2}dA+k_{5}\oint KdA,  \label{80}
\end{equation}%
where $k_{11}$ is the splay elastic constant of the Sm-A, and $k_{s}$ is
defined as $2k_{13}-k_{22}-k_{24}$, $k_{ij}$ are the Oseen-Frank elastic
constants.

The second term is the surface energy change of the inner and outer Sm-A-I
interfaces

\begin{equation}
\delta F_{A}=\gamma \oint (-2HD+d^{2}K)dA.  \label{81}
\end{equation}

The third term is the volume free energy change due to the I-Sm-A transition

\begin{equation}
\delta F_{V}=-g_{0}\oint (d-d^{2}H+\frac{1}{3}d^{3}K)dA,  \label{82}
\end{equation}%
where $g_{0}(>0)$ is the difference in the Gibbs free energy density between
Sm-A and I phase, V is the volime of Sm-A phase, and $H$ and $K$ are the
mean and Gaussian curvatures of the inner surface, respectively.

The sum of the three terms should be zero, then 
\begin{equation}
\delta F_{C}+\delta F_{A}+\delta F_{G}=0.  \label{83}
\end{equation}

Making using of the above equation and the most general equations (\ref{75}-%
\ref{79}), Prof. Ou-Yang et al. solved these equations, and gave a good
explanation of FCD in Ref. \cite{73}.

\section{Helfrich Model for Multi-layer Vesicle (II): The shapes of
fullerenes and carbon nanotubes}

Since the discovery of straight and multishell carbon nanotubes (MCTs) in
arc discharges, many unique and novel properties have been predicted for the
tubes. Ou-Yang et al. \cite{1G} analytically have obtained the general
equilibrium-shape equation of the axis curve of the MCT in the continuum
limit by taking account of competition among the curvature elasticity, the
adhesion of the interlayer van der Waals bonding, and the tension of the
outer and inner surfaces of a MCT. The sum of these three energies can be
understood as the shape formation energy.

For a single layer, the curvature elastic energy is an incremental part of
the in-layer covalent energy due to the layer curvature. Following Lenosky
et al. \cite{77}, the curvature elastic energy

\begin{eqnarray}
E_{b}^{s} &=&\varepsilon _{1}\displaystyle\sum \limits_{i}(\displaystyle%
\sum \limits_{<j>}\overrightarrow{u}_{ij})^{2}+\varepsilon _{2}%
\displaystyle\sum \limits_{<i,j>}(1-\overrightarrow{n}_{i}\cdot 
\overrightarrow{n}_{j})  \label{84} \\
&&+\varepsilon _{3}\displaystyle\sum \limits_{<i,j>}(\overrightarrow{u}%
_{i}\cdot \overrightarrow{n}_{ij})(\overrightarrow{u}_{j}\cdot 
\overrightarrow{n}_{ji}),  \notag
\end{eqnarray}%
where $u_{ij}$ is the unit vector pointing from carbon atom $i$ to its
neighbor $j$, and $\overrightarrow{n}_{i}$ is a unit vector normal to the
fullerene surface at atom $i$. The summation $\displaystyle \sum%
\nolimits_{(j)}$ is take over the three neighbor $j$ atoms to atom $i$, and
the sums of the last terms are taken over only the nearest neighbor atoms.
The first term of right hand is contribution of bond angle changes. The
second and third terms of right hand are contributions from the bending of
nearest-neighbor fullerene surface.

This discrete Lenosky's atoms interaction energy can be described using a
continuum theory of elastic energy. Then the curvature elastic energy Eq.(%
\ref{84}) of tube can be written as \cite{1G}%
\begin{equation}
E_{b}^{s}=\oint [\frac{1}{2}k_{c}(2H)^{2}+k_{G}K]dA  \label{84B}
\end{equation}%
where the bending elastic constant

\begin{equation}
k_{c}=\frac{1}{32}(18\varepsilon _{1}+24\varepsilon _{2}+9\varepsilon
_{3})(a^{2}/\sigma ),  \label{86}
\end{equation}%
\begin{equation}
k_{G}=-(8\varepsilon _{1}+2\varepsilon _{3})k_{c}/(6\varepsilon
_{1}+8\varepsilon _{2}+3\varepsilon _{3})=-1.56k_{c}.  \label{87}
\end{equation}

Let $(\varepsilon _{1},\varepsilon _{2},\varepsilon _{3})$ be value of $%
(0.96,1.29,0.05)$ eV, then $k_{c}$ is $1.17$ eV and $k_{G}$ is $-k_{c}/1.56$.

Then multi-wall fullerenes and nanotubes can be viewed as SmA LC. The shape
formation energy of the MCT can be written as 
\begin{equation}
F=F_{b}+F_{A}+F_{V},  \label{88}
\end{equation}%
where%
\begin{equation}
F_{b}={\displaystyle\sum }E_{b}^{s}=\pi (k_{c}/d)\ln (\rho _{0}/\rho _{i}),
\label{89}
\end{equation}%
\begin{equation}
F_{A}=2\pi \gamma (\rho _{0}+\rho _{i})L,  \label{90}
\end{equation}%
\begin{equation}
F_{V}=-\pi g_{0}(\rho _{0}^{2}-\rho _{i}^{2})L.  \label{91}
\end{equation}

Minimazing the shape formation energy of the MCT leads to%
\begin{eqnarray}
2\kappa _{SS}+\kappa ^{3}-2\kappa \tau ^{2}-\frac{m}{\alpha }\kappa &=&0,
\label{93} \\
\kappa ^{2}\tau &=&const,  \label{94}
\end{eqnarray}%
where $k_{ss}$=$d^{2}\mathit{k}(s)/ds^{2}$, $\mathit{k}$ are curvature, and $%
\tau $ torsion. For straight tubes, $\mathit{k,}$ $\tau =0$, while for
helical coils, $\mathit{k,}\tau =cons\tan t\neq 0$. This work has used the
same idea of the FCD formation in I and SmA LC. The relieved energy of free
carbon atoms to form graphite must be balanced by the elastic energy of
curved graphite. This work has exerted extensive academic influebce and has
been widely cited by numerous scientists \cite{78}, \cite{79}.

\section{Helfrich Variation for Reversible Transition between Peptide
Nanotobes and Spheric Vesicles Induced by Concentrating Solution}

In 2004, Stupp and colleagues \cite{80} engineered a peptide-amphiphile
molecule with distinct functional segments: a hydrophobic alkyl tail, a
peptide sequence promoting $\beta $-sheet formation, and a hydrophilic,
phosphorylated serine residue. In aqueous solution, these molecules
self-assemble into cylindrical nanofibers, presenting a dense surface of
bioactive signals. The most profound finding was that these nanofibers could
nucleate and direct the growth of hydroxyapatite crystals, the primary
mineral component of bone, along their long axis. In 2002, Zhang's group 
\cite{81} investigated a 16-residue peptide (RADA16-I) that forms stable $%
\beta $-sheet structures in salt solutions. They meticulously detailed the
hierarchical self-assembly pathway: from individual peptides to intermediate
nanofibrils, and finally to a hydrogel with a "nanofabric" morphology. The
2007 paper by Timothy J. Deming's team \cite{83} designed a simple diblock
copolypeptide consisting of polyarginine (hydrophilic, positively charged)
and polyleucine (hydrophobic). The same polyarginine segment simultaneously
plays two crucial roles. Its hydrophilicity and charge interactions,
synergizing with the hydrophobic polyleucine, drive the spontaneous
formation of stable vesicular structures in aqueous solution. Leveraging the
inherent cell-penetrating peptide (CPP) properties of arginine, these
vesicles can be efficiently internalized by various cell types and deliver
encapsulated model cargos (e.g., fluorescent dyes) into cells.

The study systematically demonstrates that for a specific peptide amphiphile
molecule, merely the physical operation of dilution can reversibly induce a
complete transition in its self-assembled structure: from rigid nanotubes $%
\rightharpoonup $ pearl-necklace intermediates $\rightharpoonup $ small and
uniform vesicles. This transition is driven by dilution altering the
molecule's critical packing parameter. As the concentration decreases, the
effective hydrophilic-hydrophobic balance of the molecules in the aqueous
environment is disrupted. To reduce interfacial energy, curvature
reorganization of the assemblies occurs: high-curvature vesicles gradually
replace low-curvature tubular structures. The "necklace-like" structure is a
crucial intermediate state in this dynamic reorganization process, providing
visual evidence for understanding the transition pathway \cite{84}.

The self-organisation of the molecules of surfactants and lipids depends on
the concentration of the lipid present in solution. Below the critical
micelle concentration the lipids form a single layer on the liquid surface
and are dispersed in solution. At the first critical micelle concentration
(CMC-I), the lipids organize in spherical micelles, at the second critical
micelle concentration (CMC-II) into elongated pipes, and at the lamellar
point (LM or CMC-III) into stacked lamellae of pipes. The CMC depends on the
chemical composition, mainly on the ratio of the head area and the tail
length.

H. Naito, M. Okuda, Z.C. Ou-Yang, regarded \cite{72}-\cite{73}
nano-structure formation in peptide as focal conic formation in SmA LC. The
focal point is what is free energy of solution-aggregate transition.

Viewing the process where dispersed monomer molecules in solution assemble
into ordered aggregates (like micelles or vesicles) as analogous to
"compressing a gas" to do work is a profound model with a clear physical
picture. The core of this analogy lies in treating the dispersed monomers in
solution as "freely moving gas molecules," while driving them into a compact
aggregate is similar to compressing the gas into a smaller volume. This
process requires overcoming entropic resistance, and thus, the work done can
be calculated.

Compressing a molecule from solution state to aggregate phase with idea gas
model need a work $k_{B}T\ln (C_{S}/C_{A})$. This work is provide by the
energy relieved due to aggregation. i.e., for an aggregate volume $\delta V$
the aggregate bulk energy is $\delta F_{V}=-g_{0}\delta V$.

Here

\begin{equation}
g_{0}=C_{A}k_{B}T\ln (C_{S}/C_{A}),  \label{95}
\end{equation}%
where $C_{A}$ is concentration of aggregate phase, $C_{S}$ is concentration
in solution.

The energy variation:

\begin{eqnarray}
F &=&\delta F_{A}+\delta F_{C}+\delta F_{V}  \label{96} \\
&=&\oint [(-\frac{1}{3}g_{0})Kd^{3}+(g_{0}H+\gamma
K)d^{2}+(2k_{11}H^{2}+k_{5}K-g_{0}-2\gamma H)d]dA.  \notag
\end{eqnarray}

Making energy variation to thickness $d$ lead to

\begin{equation}
\frac{\partial F}{\partial d}=0.  \label{97}
\end{equation}

The equation of surface is%
\begin{equation}
2k_{11}H^{2}+k_{5}K-g_{0}-2\gamma H=0.  \label{98}
\end{equation}

It is clear that the equilibrium surface must be a Weingarten surface.

For a sphere of radius $r_{0}$, we have $H=-1/r_{0}$, and $K=1/r_{0}^{2}$
and the unique solution is 
\begin{equation}
r_{0}=\frac{1}{(\gamma ^{2}+g_{0}k)^{1/2}-\gamma }.  \label{99}
\end{equation}

\begin{equation}
(\gamma ^{2}+g_{0}k)^{1/2}-\gamma \approx kg_{0}/2\gamma .  \label{100}
\end{equation}

The experimental evidence that a cylinder is a stable shape has been also
pridicated by the present theory. For a cylinder of radius $\rho _{0}$,
Ouyang et al. have $H=-1/2\rho _{0}$, and $K=0$. Then the radius of cylinder
is 
\begin{equation}
\rho _{0}=\frac{k_{11}}{(\gamma ^{2}+2k_{11}g_{0})^{1/2}-\gamma },
\label{102}
\end{equation}

\begin{equation}
(\gamma ^{2}+2k_{11}g_{0})^{1/2}-\gamma \approx k_{11}g_{0}/\gamma .
\label{103}
\end{equation}

The free energies for sphere and cylindrical tube are

\begin{equation}
F_{sphere}=-(g_{0}^{3}d^{3}/12\gamma ^{2}+g_{0}^{2}d^{2}/4\gamma ,
\label{104}
\end{equation}

\begin{equation}
F_{tube}=-g_{0}^{2}d^{2}/2\gamma .  \label{105}
\end{equation}

If $F_{tube}\geq F_{sphere}$, then $g_{0}d\geq 3\gamma $. It leads to

\begin{equation}
C_{s}\leq C_{A}e^{-3\gamma /C_{A}dk_{B}T}.  \label{108}
\end{equation}

Nanotube must transform into spherical vesicle at the Critical
Tube-to-Vesicle Transition Concentration (CTVT) during dilution where 
\begin{equation}
CTVC=CAe^{-3\gamma /C_{A}dk_{B}T}.  \label{109}
\end{equation}

Here $d$ is thickness of tube, and $\gamma $ is tension of solution/
aggregate interface \cite{1H}. It can be concluded that nanotube must
transform into spherical vesicle at the Critical Tube-to-Vesicle Transition
Concentration during dilution.

If $k_{5}$ value is ignored in Eq. (\ref{98}), then it is obtained that

\begin{equation}
2k_{11}H^{2}-g_{0}-2\gamma H=0.  \label{110}
\end{equation}

The solution of the above shape equation is a surface with $H$=constant. In
1884, Delaunay \cite{86} find beautiful way to construct such a surface with
rotationally symmetry: By rolling a given conic section on a straight line
in a plane, and then rotating the trace of a focus about the line, one
obtains the surface.

The nanotube to spherical vesicle transition is linked through joined
necklace-like structures. The metastable necklace-like structure can be
described as Delaunay surface. The theory of Delaunay surfaces can predict
geometric parameters of the necklace-like structure, such as its period, and
the ratio of bulge size to neck diameter. These predictions can be
quantitatively compared with experimental measurements.

The beautiful mathmatical concept of "constant mean curvature" connects
seemingly different stable states (tube, sphere) and metastable states
(necklace), which shows they are different stationary points on the same
energy landscape.

\section{Helfrich Model for The Elastic Energy Model of Icosahedral
Self-Assemblies}

Experimentally, the capsids of many viruses are observed to exhibit
icosahedral symmetry. Prominent examples include the Cowpea Chlorotic Mottle
Virus (CCMV), whose structure determined by cryogenic-temperature
transmission electron microscopy (cryo-TEM) is a truncated icosahedron \cite%
{87}, and the Herpes Simplex Virus, which is approximately five times larger
than CCMV but also displays an icosahedral architecture as revealed by
cryo-TEM reconstruction \cite{88}.

Interestingly, this structural preference extends beyond biological systems
into purely physicochemical assemblies. Recently, surfactant aggregates have
also been found to form icosahedral bilayers \ref{89}. This convergence
raises a fundamental question: why is the icosahedron, among the five
Platonic solids, the preferred geometry for such closed-shell structures in
both biology and soft matter?

Early theoretical attempts to explain viral architecture, such as the
geometrical "parking" model proposed by Crick and Watson, focused on the
principle that a small virus must be constructed from the regular
aggregation of many identical, asymmetric protein subunits \cite{90}. While
foundational, this model predicted structures with cubic symmetry, which
conflicted with the predominant icosahedral symmetry observed in nature. A
more significant discrepancy was that the number of morphological units
(capsomeres) on the surface of actual icosahedral viruses is typically
greater than 60 and often not a simple multiple of 60, contrary to the
model's basic prediction.

The resolution to this paradox was provided by the quasi-equivalence theory
developed by Caspar and Klug. This seminal theory elegantly explains how
more than 60 chemically identical subunits can assemble into an icosahedral
shell by allowing for slight, permissible distortions in their bonding
interactions, thereby accommodating the curvature necessary to form a closed
container while maximizing genetic economy and structural stability.

On the Quasi-Equivalence Theory and Its Puzzles: From Structural Prediction
to Energetic Limitations

The quasi-equivalence theory, formulated by Caspar and Klug, provides a
fundamental framework for understanding the architecture of icosahedral
viral capsids. The core premise is that the capsid surface can be subdivided
into a network of quasi-equivalent triangles, which are organized into
clusters of hexamers and pentamers. This is conceptually represented by
subdividing each triangular face of a base icosahedron into smaller,
congruent triangular facets---for instance, into nine facets per
face---thereby allowing the assembly of a shell from a larger number of
identical protein subunits while maintaining overall icosahedral symmetry 
\cite{91}.

However, this elegant theory faces empirical puzzles. For viruses such as
Polyoma virus \cite{92} and Simian Virus 40 \cite{93}, the Caspar-Klug model
predicts a capsid built from a combination of hexagonal and pentagonal
capsomeres. Contrary to this prediction, high-resolution structural studies
revealed that these viruses are constructed exclusively from pentagonal
morphological units.

This discrepancy highlights a significant limitation in the prevailing
approaches at the time. Theories like the quasi-equivalence model primarily
focused on the geometric packing and symmetry of molecular subunits within
the capsid. They largely overlooked a critical physical factor: the elastic
energy of the capsid shell itself. The mechanical properties and bending
costs associated with deforming protein subunits into a closed, curved
surface are now understood to be crucial for determining final assembly
morphology, explaining why simple geometric packing rules sometimes fail to
predict observed structures.

Our model will estimate the elastic energy of icosahedron based on Lenosky's
lattice model which describes the energy of curved graphite layers

\begin{equation}
E_{b}=\varepsilon _{1}\displaystyle\sum \limits_{i}(\displaystyle\sum %
\limits_{\langle j\rangle }\overrightarrow{u}_{ij})^{2}+\varepsilon _{2}%
\displaystyle\sum \limits_{\langle i,j\rangle }(1-\overrightarrow{n}%
_{i}\cdot \overrightarrow{n}_{j})+\varepsilon _{3}\displaystyle\sum %
\limits_{\langle i,j\rangle }(\overrightarrow{n}_{i}\cdot \overrightarrow{u}%
_{i,j})(\overrightarrow{n}_{j}\cdot \overrightarrow{u}_{j,i}).  \label{112}
\end{equation}

Ou-Yang et al. \cite{1G} have derived the continuum form of the Helfich
elastic energy for a smoothly curved fullerene from Lenosky's lattice model.
Making using the same idea, the continuum form of the Helfich elastic energy
for a polyhedra can be obtained also. If subunit and its nearest neighbors
are in the same plane, its contributions to the elastic energy is zero, only
the subunits besides each edge contribute to the elastic energy. Then the
continuum form of the Helfich elastic energy for a polyhedra is \cite{94}

\begin{equation}
E_{bc}^{(s)}=N\{\varepsilon _{1}(4-4\cos \beta )+\varepsilon _{2}[4(1-\frac{%
\cos \beta +1/2}{\sqrt{5/4+\cos \beta }}+\frac{\sin ^{2}\beta }{2(5/4+\cos
\beta )})]+\varepsilon _{3}\frac{\sin ^{2}\beta }{4(5/4+\cos \beta )}\}.
\label{115}
\end{equation}

Here%
\begin{equation}
\beta =\pi /2-\alpha /2,  \label{116}
\end{equation}
\begin{equation}
N=\frac{\sqrt{S}}{\alpha \sqrt{2-2\cos \theta }}\sqrt{2N_{1}\tan (\pi /p)},
\label{117}
\end{equation}%
where $N$ is molecular number in edges related to the surface area and kind
of polyhedra.

When comparing the icosahedron with the other Platonic solids, it is found
to possess higher symmetry. This conclusion aligns with the analysis
presented in the comparison of the Icosahedron and the Other Four Regular
Polyhedra (Tetrahedron, Cube, Octahedron, Dodecahedron). This provides
subunits with the most diverse, energetically similar binding orientations,
favoring error correction and robust assembly. Lower symmetry, offering
fewer equivalent binding orientations. The assembly pathway has poor fault
tolerance. It is found that the elastic energy of icosahedron is the lowest
and the energy increases with area.

The Helfrich elastic energy of a sphere \cite{1G}%
\begin{equation}
E_{bs}^{(s)}=\oint [\frac{1}{2}k_{c}(2H)^{2}+\overline{k}K]dA=4\pi (2k_{c}+%
\overline{K}),  \label{118}
\end{equation}

where%
\begin{equation}
k_{c}=2\varepsilon _{1}+10\varepsilon _{2}/9+\varepsilon _{3}/9,  \label{119}
\end{equation}%
\begin{equation}
\frac{\overline{k}}{k_{c}}=\nu -1.  \label{120}
\end{equation}

Then, the elastic energy of a sphere is%
\begin{equation}
E_{bs}^{(s)}=4\pi (1+\nu )(2\varepsilon _{1}+10\varepsilon
_{2}/9+\varepsilon _{3}/9).  \label{121}
\end{equation}

According to a July 19th, 2013 report by the British newspaper The Daily
Mail, French scientists made a significant discovery: a new,
never-before-seen giant virus. Dubbed Pandoravirus, its remarkable size of
approximately one micrometre---about ten times larger than typical
viruses---and unusual characteristics led to sensational speculation about a
potential extraterrestrial origin, such as Mars.

This discovery highlights a category of "giant viruses." For instance, the
Megavirus, another large virus, exhibits a polyhedral structure and measures
up to 440 nm in length, ranking among the largest known on Earth.

The structural analysis of such complex biological assemblies, including
viral capsids and membranes, finds a relevant theoretical framework in the
book "Geometric Methods in the Elastic Theory of Membrane in Liquid Crystal
Phases" by Ou-Yang Zhong-Can, Liu Ji-Xing, and Xie Yu-Zhang \cite{95},
published by World Scientific in 1999. The principles discussed therein can
provide insights into the physical constraints and geometric organization
governing these massive viral structures.

In the "Introduction of New Books" section of the Journal of the Physical
Society of Japan (Vol. 55, No. 4, 2000), Ryohta Morikawa from Tokyo
University of Pharmacy and Life Sciences highlighted the textbook "Elastic
Theory of Membrane in Liquid Crystal Phases" \cite{95}". The work presents a
systematic treatment of membrane elastic mechanics grounded in the
principles of liquid crystal physics.

For recent theoretical advances in elasticity of membranes following
Helfrich's spontaneous curvature model, please refer to Reference \cite{96}.

The textbook "Molecular and Cell Biophysics" \cite{97} regards W. Helfrich
LC curvature elasticity model of membranes as the interpretation of RBC
shape. Professor Zhongcan Ouyang was honored as one of the Recipients of the
2015 (9th) Japan Society of Applied Physics (JSAP) International Fellow
Award for pionering research on material shapes.

\section{Shape formation in 2D lipid Monolayer at air/water interface}

The Brewster Angle Microscopy (BAM) measurement system enables the direct
observation of lipid monolayers at the air-water interface. This technique
exploits the fact that different lipid phases exhibit distinct reflectivity
under Brewster angle illumination, allowing for the visualization and
imaging of phase domains within the monolayer.

A key application is the study of shape formation in two-dimensional lipid
domains. Notably, these domains often deviate from simple, compact circular
shapes. A common observation is the appearance of cusps (sharp points) at
the domain boundaries. These give rise to distinctive, non-circular
morphologies such as "Boojum-like" (or "Peach-like") and "Kidney-like"
domains. The term "Boojum" is borrowed from Lewis Carroll's poem The Hunting
of the Snark, where it describes a mythical and perilous creature, here
metaphorically applied to these complex, cusped shapes.

Furthermore, under specific conditions, the equilibrium shape may not be a
disk but can instead form a torus (a doughnut-like ring). These
morphological studies are detailed in Ref. \cite{98}.

An approximate theoretical framework \cite{99}-\cite{100} for describing
shape formation in two-dimensional lipid domains often necessitates the
introduction of an artificial cut-off in the calculation. This is typically
required to handle the long-range nature of the electrostatic interactions
within the monolayer. The theory places significant emphasis on the
contribution of the energy arising from dipole-dipole interactions (D) to
the total free energy of a domain, which is a key factor in determining its
equilibrium shape.

The energy of dipole-dipole interaction \cite{100}.

\begin{equation}
F_{dipole}=-\int \frac{\overrightarrow{\mu }(\overrightarrow{r})\cdot 
\overrightarrow{\mu }(\overrightarrow{r^{\prime }})}{|\overrightarrow{r}-%
\overrightarrow{r^{\prime }}|^{3}}dA=\frac{-\mu ^{2}}{2}\oint \oint \frac{%
\overrightarrow{t}(l)\cdot \overrightarrow{t}(s)}{|\overrightarrow{r}(l)-%
\overrightarrow{r}(s)|}dlds,  \label{122}
\end{equation}%
where $\mu $ is dipole density, 
\begin{equation}
\overrightarrow{t}(s)=d\overrightarrow{r}(s)/ds,  \label{123}
\end{equation}%
and $l$, $s$ are contour lengths.

Considering the Helfrich free energy, the total free energy can be written
as \cite{101} 
\begin{equation}
F=\Delta P\int dA+\gamma \oint ds-\frac{1}{2}\mu ^{2}\oint \oint \frac{%
\overrightarrow{t}(l)\cdot \overrightarrow{t}(s)}{|\overrightarrow{r}(l)-%
\overrightarrow{r}(s)|}dlds.  \label{124}
\end{equation}

Here $\overrightarrow{t}(s)$ is tangential vector of boundary of domain. $%
\gamma $ is line tension. $\Delta P$ is a Lagrange multiplier and 
\begin{equation}
\Delta P=-g_{0},  \label{125}
\end{equation}%
where $g_{0}$ is Gibbs free energy density difference between outer (fluid)
and inner (solid) phases.

A key step is to rewitten the dipolar force energy as

\begin{equation}
F_{\mu }=-\frac{1}{2}\mu ^{2}\oint (\oint \frac{\overrightarrow{t}(s)\cdot 
\overrightarrow{t}(s+x)}{|\overrightarrow{r}(s)-\overrightarrow{r}(s+x)|}%
dx)ds,  \label{126}
\end{equation}%
where arc-variable $x$ is defined as $x\equiv l-s$.

Let $\kappa $ is curvature of boundary. Based on Frenet formulas of a plane
curve, one has%
\begin{equation}
d\overrightarrow{t}/ds=\kappa \overrightarrow{m}(s),  \label{127}
\end{equation}

\begin{equation}
d\overrightarrow{m}/ds=-\kappa (s)\overrightarrow{t}(s).  \label{128}
\end{equation}

Introducing artificial cutoff to prevent divergence:

\begin{equation}
|\overrightarrow{r}(s+x)-\overrightarrow{r}(s)|\geq h.  \label{129}
\end{equation}

One can obtain%
\begin{equation}
\overrightarrow{t}(s+x)=\overrightarrow{t}(s)+\kappa (s)\overrightarrow{m}%
(s)x+\frac{1}{2}[\kappa (s)\overrightarrow{m}(s)-\kappa ^{2}(s)%
\overrightarrow{t}(s)]x^{2}...,  \label{130}
\end{equation}

\begin{equation}
\overrightarrow{r}(s+x)=\overrightarrow{r}(s)+\overrightarrow{t}(s)x+\frac{1%
}{2}\kappa (s)\overrightarrow{m}(s)x^{2}...,  \label{131}
\end{equation}%
Then it is obtained that%
\begin{equation}
F_{\mu }=-\frac{1}{2}\mu ^{2}\ln \frac{L}{h}\oint ds+\frac{11}{96}\mu
^{2}L^{2}\oint^{2}\kappa (s)ds,  \label{132}
\end{equation}%
The above equation is just equation of negative line tension and bending
energy. $L$ is boundary length, integral $dx$ from $0$ to $L$ induces
divergence, so integral takes from $h$-(cutoff ) to $L$, $h$ is artificially
regarded as dipole-dipole separation distance, or the thickness of the
monolayer.

Making use of the above equation, the Eq. (\ref{124}) becomes%
\begin{equation}
F=\Delta P\int dA+\lambda \oint ds+\alpha \oint \kappa ^{2}ds,
\label{133}
\end{equation}%
where 
\begin{equation}
\lambda =\gamma -\frac{\mu ^{2}}{2}\ln \frac{L}{h},\qquad \alpha =\frac{11}{%
96}\mu ^{2}L.  \label{134}
\end{equation}

The positive $\lambda $ will shorten boundary, and circle-like domain is
given. The negative $\lambda $ will thin the shapes with increasing size,
and the circle is instable.

Let $\delta F=0$, then one obtains%
\begin{equation}
\Delta P-\lambda \kappa +\alpha \kappa ^{3}+2\alpha \kappa _{SS}=0,
\label{135}
\end{equation}%
where%
\begin{equation}
\kappa _{SS}=d^{2}\kappa /ds^{2},  \label{136}
\end{equation}%
and%
\begin{equation}
\lambda =\gamma -\frac{\mu ^{2}}{2}\ln \frac{L}{he}+\frac{11}{48}L\mu
^{2}\oint \kappa ^{2}ds.  \label{137}
\end{equation}%
Here $\lambda $ is line tension depended on shape and size.

A 2D circle of radious $\rho _{0}$ is a cubic curve. If inner phase is
solid, then 
\begin{equation}
\kappa =-1/\rho _{0}.  \label{138}
\end{equation}

Else if inner phase is fluid, then%
\begin{equation}
\kappa =1/\rho _{0}.  \label{139}
\end{equation}

Shape equation of circle domain becomes%
\begin{equation}
\Delta P(\kappa )=\lambda \kappa -\alpha \kappa ^{3},\qquad (\alpha \geq 0).
\label{140}
\end{equation}

This theory has predicated many types of solution such as two circles, two
tori, only one circle and no compact circular domain for $\lambda ,\Delta
P\leq 0$ and so on.

The quasi-polygon domain is seen as a branching phenomenon of the
instability of a circle of radius $\rho _{0}$. Let the circle has a slightly
distorsion, the 
\begin{equation}
\rho =\rho _{0}+\displaystyle \sum\nolimits_{m}b_{m}\exp (im\phi ),
\label{141A}
\end{equation}%
where $i\cdot i=-1$, $0\leq \phi \leq 2\pi $, $m=0$, $\pm 1,\pm 2,...,\pm
\infty $.

If the 2D domain has not dipole tilt, i.e. $\mu _{\parallel }=0$, Iwamoto,
Fei Liu, Z.C. Ou-Yang have pridicated an $m$-th harmonic shape transition
under the following critical surface pressure \cite{102}

\begin{equation}
\Delta P=\frac{11\pi ^{2}\mu _{\perp }^{2}(m^{2}-1)}{12\rho _{0}},
\label{141}
\end{equation}

If $\Delta P=0$, then analytical solution of $r(s)=(x(s),y(s))$ as follows%
\begin{equation}
\widetilde{x}(\widetilde{s})=-2\int\nolimits_{0}^{\widetilde{s}^{\prime
}}sn(s^{\prime })cn(s^{\prime })ds^{\prime },  \label{142}
\end{equation}

\begin{equation}
\widetilde{y}(\widetilde{s})=\int\nolimits_{0}^{\widetilde{s}^{\prime
}}[1-2sn^{2}(s^{\prime })ds^{\prime }.  \label{143}
\end{equation}

If $\Delta P\neq 0$, with $\kappa _{s}$ and integrating, its first integral
is obtained%
\begin{equation}
\Delta P_{\kappa }-\frac{\Lambda }{2}\kappa ^{2}+\frac{\alpha }{4}\kappa
^{4}+\alpha \kappa _{s}^{2}=C,  \label{144}
\end{equation}%
\begin{equation}
\kappa (s)-\kappa (0)=\frac{6(\overline{\Lambda }\kappa (0)-\Delta \overline{%
P}-\kappa ^{3}(0)}{24\wp (s)+3\kappa ^{2}(0)-\Lambda },  \label{145}
\end{equation}%
where%
\begin{equation}
\wp (s)=e_{3}+\frac{e_{1}-e_{3}}{sn^{2}(\sqrt{e_{1}-e_{3}}\widetilde{s})}.
\label{146}
\end{equation}

In practical calculations the Weierstrass elliptic funtion $\wp (s)$ has to
be converted into the forms in common use, e.g., Jacobi's elliptic functions.

Through rigorous geometric variational calculus, singularity analysis, and
numerical computation, The work of Ou-Yang et al. transformed the early
physical insights of Andelman et al. into a complete mathematical model
amenable to quantitative computation and prediction. Its core mathematical
formulation---reformulating the long-range dipole-dipole force as a linear
functional of the boundary curvature---provided a powerful analytical tool
for this class of problems. The methodology has also influenced subsequent
research on topics such as the assembly of colloidal particles at interfaces
and vesicle morphology.

\section{Mathematical structure of axisymmetric biomembrane shape equation}

\subsection{Mathematic basis of prolonged Lie derivative on biomembrane
shape equation}

The quest to understand and predict the equilibrium shapes of
interfaces---from soap films and liquid droplets to red blood cells and
crystalline solids---constitutes a cornerstone of materials science and soft
matter physics. The modern membrane shape equation is the culmination of
centuries of thought, elegantly synthesizing principles from geometry,
thermodynamics, and elasticity. The microsopic structure of a material
determines its macroscopic geometric features and the governing shape
equations, which can be characterized by symmetries.

The group structure is obtained \cite{104} for axisymmetric membrane shape
equation by Chern's $3$-order differential equation theory \cite{105}. The
shape equation of rotationally symmetric vesicles is $3$-order differential
equation (\ref{28}). Based on Chern's $3$-order differential equation
theory, the invariant of axisymmetric shape equation can be calculated. One
may find that the relative invariance does not vanish, it is also possible
to define a generalized geometry in the plane with the elements of contact
of the second order $x$, $y$, $y^{\prime }$, $y"$\ as the elements of the
space and with a certain five-parameter group as its fundamental group. In
the example of axisymmetric membrane shape equation, one may find that the
membrane shape is a five-parameter group and characterized by twelve group
structure parameters which are functions of pressure difference, tensile
stress and asymmetry effect of the membrane or its environment. When these
varieties of membrane or environment change, the structure constants vary;
then one can obtain directly the change of symmetric group and the
information on the membrane shape variation.

In addition, a group analysis of the axisymmetric membrane shape equation
based on metric tensor has been given by conformal form transformation \cite%
{106}-\cite{108}.

This section will employ the extended Lie group method to systematically
analyze the mathematical structure of the axisymmetric Zhong-Can-Helfrich equation and classify their analytical solutions. Nonanalytic solutions can
be viewed as topological deformations of these analytic ones. The analytic
solutions thus serve as anchoring points within the vast sea of non-analytic
solutions.

In fact, the $3$-order shape equation (\ref{28}) can be varied to the $2$%
-order equation (\ref{33}) by the first integration. Then obtaing the
mathematic structure become more easier. Here we will discuss the group
structure of the $2$-order differential equation by pronged Lie group
operator \cite{109}.

The biomembrane shape Eq. (\ref{33}) can be written as%
\begin{equation}
\rho (\Psi ^{2}-1)\Psi ^{\prime \prime }-\frac{\Phi \rho }{2}(\Psi ^{\prime
})^{2}+(\Psi ^{2}-1)\Psi ^{\prime }+\frac{\Psi }{\rho }-\frac{\Psi ^{3}}{%
2\rho }-c_{0}\Psi ^{2}+\overline{\lambda }\rho \Psi +\frac{\overline{p}}{2}%
\rho ^{2}+C_{3}=0\text{,}  \label{147}
\end{equation}%
where $C_{3}$ is an integral constant.

The infinitesimal generator is a first-order linear differential operators%
\begin{equation}
\mathbf{X=\xi }(\rho ,\Psi )\frac{\partial }{\partial \rho }+\eta (\rho
,\Psi )\frac{\partial }{\partial \Psi }  \label{148}
\end{equation}%
The membrane shape equation admit a one-parameter group with the
infinitesimal generator $\mathbf{X}$ if and only if the following
infinitesimal condition holds%
\begin{equation}
pr^{(2)}\mathbf{X}[\Psi (\rho ,\Psi )]=0,  \label{149}
\end{equation}%
where $pr^{(2)}\mathbf{X}$ is the second-order prolongation of $\mathbf{X}$%
\begin{equation}
pr^{(2)}\mathbf{X=\mathbf{\xi }(\rho ,\Psi )\frac{\partial }{\partial \rho }%
+\eta (\rho ,\Psi )\frac{\partial }{\partial \mathbf{\Psi }}+}\eta _{\lbrack
1]}\frac{\partial }{\partial \mathbf{\Psi }_{\rho }}+\eta _{\lbrack 2]}\frac{%
\partial }{\partial \mathbf{\Psi }_{\rho \rho }}.  \label{150}
\end{equation}%
Here 
\begin{eqnarray}
\mathbf{\eta }_{[1]} &=&D_{\rho }\eta -\mathbf{\Psi }_{\rho }D_{\rho }\xi
,\qquad D_{\rho }=\frac{\partial }{\partial \rho }+\mathbf{\Psi }_{\rho }%
\frac{\partial }{\partial \mathbf{\Psi }}  \label{151} \\
\eta _{\lbrack 2]} &=&D_{\rho }\eta _{\rho }-\mathbf{\Psi }_{\rho \rho
}D_{\rho }\xi -\mathbf{\Psi }_{\rho \phi }D_{\rho }\eta .  \label{152}
\end{eqnarray}

\subsection{The Lie operator of sphere solution}

In the case of sphere solution,

\begin{equation}
\Psi ^{^{\prime }}=C.  \label{153}
\end{equation}

Then $\Psi ^{^{\prime \prime }}=0$. The membrane shape equation (\ref{147})
is simplified as%
\begin{equation}
-\frac{\Psi \rho }{2}C^{2}+(\Psi ^{2}-1)C+\frac{\Psi }{\rho }-\frac{\Psi ^{3}%
}{2\rho }-C_{0}^{2}\Psi +\widetilde{\lambda }\rho \Psi +\frac{\widetilde{p}}{%
2}\rho ^{2}+C_{3}=0.  \label{154}
\end{equation}

Then the operator can be also simplified as%
\begin{equation}
\xi =a_{1}+a_{2}\rho +a_{3}\Psi ,\qquad \eta =b_{1}+b_{2}\rho +b_{3}\Psi .
\label{155}
\end{equation}

Substitute the above operator into equation (\ref{154}), one obtains%
\begin{equation}
-\frac{C^{2}}{2}\xi \Psi -\frac{\Psi \xi }{\rho ^{2}}+\frac{\Psi ^{3}\xi }{%
2\rho ^{2}}+\widetilde{\lambda }\Psi \xi +\widetilde{p}\rho \xi -\frac{%
C^{2}\eta \rho }{2}+2C\eta \Psi +\frac{\eta }{\rho }-\frac{3\Psi ^{2}\eta }{%
2\rho }-2C_{0}\Psi \eta +\widetilde{\lambda }\rho \eta =0.  \label{157}
\end{equation}

Using equation (\ref{150}, \ref{153}), one can obtain%
\begin{equation}
b_{2}=0,\qquad b_{3}=a_{2},\qquad a_{3}=0.  \label{158}
\end{equation}

Taking use of Eq. (\ref{153}), the function $\Phi $ can be written as%
\begin{equation}
\Psi =C\rho +C_{4},  \label{159}
\end{equation}%
Substituting the results of Eqs. (\ref{155}, \ref{158}, \ref{159}) into Eq. (%
\ref{157}), one can obtain $C_{4}=0$, or $C_{4}^{2}=2$.

For the case $C_{4}=0$, it is obtained that 
\begin{equation}
b_{1}=a_{1}C,  \label{160}
\end{equation}

\begin{equation}
a_{2}=0,\qquad or\qquad \widetilde{p}=2c_{0}C^{2}-2\widetilde{\lambda }C.
\label{161}
\end{equation}

The operator of sphere solution becomes%
\begin{equation}
\xi =a_{1}+a_{2}\rho ,\qquad \eta =a_{1}C+a_{2}\Psi ,  \label{162}
\end{equation}%
which defines a group $g_{1}$ for spherical membrane shape.

Then operator is 
\begin{equation}
\mathbf{X=}a_{1}(\frac{\partial }{\partial \rho }+C_{2}\frac{\partial }{%
\partial \Psi })+a_{2}(\rho \frac{\partial }{\partial \rho }+\Psi \frac{%
\partial }{\partial \Psi }),  \label{163}
\end{equation}%
which is composed by two basis%
\begin{eqnarray}
\mathbf{X}_{1} &=&\frac{\partial }{\partial \rho }+C\frac{\partial }{%
\partial \Psi },  \label{164} \\
\mathbf{X}_{2} &=&\rho \frac{\partial }{\partial \rho }+\Psi \frac{\partial 
}{\partial \Psi }.  \label{165}
\end{eqnarray}

Its first integral invariant of $\mathbf{X}_{1}$ is%
\begin{equation}
\Phi -C\rho =c_{s1},  \label{166}
\end{equation}%
where $c_{s1}$ is a constant. The case of $c_{s1}=0$ is corresponding to the
nontrival sphere solution, and $C$ is curvature of sphere.

Its first integral invariant of $\mathbf{X}_{2}$ is%
\begin{equation}
\Phi -c_{s2}\rho =0,  \label{167}
\end{equation}%
where $c_{s2}$($\gtrdot 0$) is a constant. This case is corresponding to the
trival sphere solution or nontrival sphere solution.

The commutator of the operators $\mathbf{X}_{1}$ $\mathbf{X}_{2}$ for Eqs. (%
\ref{164},\ref{165}) is 
\begin{equation}
\lbrack \mathbf{X}_{1},\mathbf{X}_{2}]=\mathbf{X}_{1}.  \label{168}
\end{equation}

For the case $C_{4}=\pm \sqrt{2}$, it is obtained that%
\begin{equation}
\xi =a_{2}(\pm \frac{\sqrt{2}}{C}+\rho ),\qquad \eta =a_{2}\phi .
\label{169}
\end{equation}

Let small radius of Clifford torus $r=1/C$, and noticing that $|\Psi |\leq 1$%
,$\ $then the Clifford torus is written as%
\begin{equation}
\Psi =\frac{\rho }{r}-\sqrt{2}.  \label{169B}
\end{equation}

The operator (\ref{169}) becomes 
\begin{equation}
\mathbf{X}=(-\frac{\sqrt{2}}{C}+\rho )\frac{\partial }{\partial \rho }+\Psi 
\frac{\partial }{\partial \Psi },  \label{170}
\end{equation}%
which is a translation of circle along $\rho $ direction, and the
translation distance is $\frac{\sqrt{2}}{C}$.

\subsection{The Lie operator of biconcave discoid and Delaunay curve}

Let $\Psi =C_{0}\rho \ln \rho +C\rho $, which is corresponding to the
biconcave discoid shape of red blood shape. Making the same method, the
shape biconcave discoid has operators%
\begin{equation}
\mathbf{X}=\rho \frac{\partial }{\partial \rho }+(C_{0}\rho +\Psi )\frac{%
\partial }{\partial \Psi }.  \label{171}
\end{equation}

The operator of cylinder is%
\begin{equation}
\mathbf{X}=\frac{\partial }{\partial \rho },  \label{172A}
\end{equation}%
or%
\begin{equation}
\mathbf{X}=\frac{\partial }{\partial \Psi }  \label{172B}
\end{equation}

Let $\Psi =\frac{C_{1}}{\rho }+\frac{C_{0}}{2}\rho $, which is Delaunay
curve. The operators of Delaunay surface is

\begin{equation}
\mathbf{X}=\rho \frac{\partial }{\partial \rho }+(C_{0}\rho -\Psi )\frac{%
\partial }{\partial \Psi }.  \label{173A}
\end{equation}

In addition, the Delaunay has a extened form \cite{110} 
\begin{equation}
\Psi =\frac{1}{C_{0}\rho }+\frac{C_{2}}{2}\rho +C_{6},  \label{173B}
\end{equation}%
whose generator is%
\begin{equation}
\mathbf{X}=\rho \frac{\partial }{\partial \rho }+(C_{2}\rho +C_{6}-\Psi )%
\frac{\partial }{\partial \Psi }  \label{174}
\end{equation}

From Eqs. (\ref{171})-(\ref{173A}), we can conclude the Lie operatos of
sphere, biconcave discoid, Delaunay surface and cylinder:%
\begin{eqnarray}
\mathbf{X}_{1} &=&\frac{\partial }{\partial \rho },  \label{176} \\
\mathbf{X}_{2} &=&\frac{\partial }{\partial \Psi },  \notag \\
\mathbf{X}_{3} &=&\rho \frac{\partial }{\partial \rho }+(C_{2}\rho +\Psi )%
\frac{\partial }{\partial \Psi },  \notag \\
\mathbf{X}_{4} &=&\rho \frac{\partial }{\partial \rho }+(C_{2}\rho -\Psi )%
\frac{\partial }{\partial \Psi },  \notag \\
\mathbf{X}_{5} &=&(\rho +C_{3})\frac{\partial }{\partial \rho }+\Psi \frac{%
\partial }{\partial \Psi },  \notag \\
\mathbf{X}_{6} &=&\rho \frac{\partial }{\partial \rho }+(C_{2}\rho
+C_{6}-\Psi )\frac{\partial }{\partial \Psi }.  \notag
\end{eqnarray}%
Here $\mathbf{X}_{1}$, $\mathbf{X}_{2}$ are the generators of cylinder, $%
X_{3}$ is generator of biconcave discoid, $X_{4}$ and $X_{6}$ are generator
of Delaunay surfaces, $X_{5}$ is generator of sphere or Clifford torus.
These $6$ generators form a group by the commutators of operators in Table $%
1 $.

The cylinder satisifies translator transformation, so it can form group with
sphere and Clifford torus, whose communitators can be seen in Table $2$.

A synthesis of Eqs. (\ref{163})-(\ref{176}), it can be conclude that the
mathematical structure of axisymmetric membran shape equation, with $6$
foudmantal shape with $5$ basis generators \{$\frac{\partial }{\partial \rho 
},\frac{\partial }{\partial \Psi },\rho \frac{\partial }{\partial \rho }%
+\Psi \frac{\partial }{\partial \Psi },\rho \frac{\partial }{\partial \rho }%
-\Psi \frac{\partial }{\partial \Psi },\rho \frac{\partial }{\partial \Psi }$%
\}, and $8$ structure parameters, forms a $6$-parameter group. This result
is the extension and application of $3$-order differential equation of Prof.
Chern's result \cite{105}. Additionally, it should be emphasized here that
shapes such as cylinders, spheres, tori, bicocave discoids and Delaunay
surfaces form groups. This result is merely a geometric feature of the
shapes and is independent of the membrane equation. When the pressure on the
membrane, surface tension, and bending modules meet certain conditions, the
biomembrane will take on the aforementional types of shapes.

The first integral invariant of the Lie derivative for the membrane shape
equation correponds to the solutions of the equation. In this way, all
analytic solutions of the membrane shape equation can be put in one-to one
correspondence with and characterizeed by the Lie derivatives. Various
membrane shapes posses their own unique Lie derivatives. The commutation
relations between these Lie derivatives and the various groups they form
reveal the underlying connections through which these shapes interact and
transform into one another.

\section{Conclusion}

The theoretical framework for understanding complex structures in soft
matter, such as the focal conic domains in liquid crystals, was established
through a foundational sequence of work by Professor Wolfgang Helfrich and
profoundly extended and applied by Professor Ou-Yang.

Professor Helfrich's seminal contribution was the establishment of a
universal continuum theory in his 1973 paper. He introduced the Helfrich
free energy which describes the elastic energy of a thin film based purely
on its geometry---specifically, its mean curvature and Gaussian curvature.
This model abstracted complex microscopic interactions into an elegant
geometric language, creating a unified framework applicable to biological
lipid bilayers and, crucially, lamellar liquid crystal phases. It provided
the energetic principle---minimization of curvature elastic energy---that
would answer "why" certain shapes form.

Professor Ou-Yang's pivotal work was to solve, generalize, and apply
Helfrich's framework, transforming it into a powerful tool for predicting
specific shapes across physics and biology. In his celebrated collaboration
with Helfrich (1987, 1989), Professor Ou-Yang employed rigorous variational
calculus to derive the general shape equation from the Helfrich free
energy---the "Zhong-Can-Helfrich equation." This allowed for the first
quantitative predictions of complex vesicle shapes, such as the biconcave
disc of red blood cells and toroidal vesicles.

Professor Ou-Yang recognized the universal power of this theory. He
demonstrated that the incompressibility constraint in smectic A liquid
crystals is mathematically isomorphic to that in membranes. By solving the
shape equation for this case, he proved that the energy-minimizing structure
is the Dupin cyclide, thereby providing the complete energetic justification
for the focal conic domains observed microscopically. This directly
connected Helfrich's biomembrane theory to the core problem of liquid
crystal defects. He extended this curvature elasticity approach far beyond
its origins, applying it to explain the mechanics of carbon nanotubes (via a
continuum limit of discrete models) and the assembly of viral capsids. His
work established that the Zhong-Can-Helfrich formalism is the "starting point"
for analyzing shape and stability in virtually any soft matter system
governed by bending energy, from self-assembled peptides to two-dimensional
materials.

In summary, Professor Helfrich provided the fundamental language of
curvature elasticity, while Professor Ou-Yang mastered its grammar and wrote
its most important chapters across multiple disciplines. Together, their
work shows that the morphology of soft matter---from living cells to
synthetic materials---is a manifestation of geometry optimized under energy
constraints, a cornerstone concept in modern physics.

This review article summarizes Prof. Ou-Yang's numerous outstanding
contributions to the science of soft matter. It highlights the fundamental
approach of continuously bridging scales: from the microscopic structure and
intermolecular forces of a material to a macroscopic free energy; from the
minimization of this free energy to governing macroscopic shape equations;
and from these differential shape equations to the resulting geometric
structures. It demonstrates how the microscopic architecture of a material
dictates the geometric symmetry of its macroscopic form, thereby revealing
the profound connection between the microscopic constitution
of a material, its structure and interactions, and its ultimate macroscopic geometry.

At the end of this article, we express our profound mourning for the great scientist, Prof. Podgornik. Professor Podgornik was a master at deep understanding the physical world. Throughout his brilliant academic career, he delved into research areas such as electrified fluids, Lifshitz dispersion interaction theory, Casimir effect, and biological membranes. In particular, he made pioneering contributions to the study of electrostatic interactions in polyelectrolytes, van der Waals forces, and the Casimir effect between macromolecules as well as to the physical problems of DNA and viruses, establishing himself as a benchmark and pioneer in these fields. His work style, characterized by rigorous theoretical and computational analysis, demonstrated extraordinary depth and breadth. Although Professor Podgornik has left us, his academic achievements and scientific spirit will always inspire us to forge ahead on the path of scientific research. 

\textbf{Data Availability statement}

All data underlying the results are available as part of the article and no
additional source data are required.

\textbf{Conflict of interest}

The authors of this work declare that they have no Conflict of interest.

\textbf{Acknowledgements}

The author Tao Xu is grateful to Prof. Zhong-Can Ou-Yang for his
supervision. The author thanks Prof. Haijun Zhou, Prof. Tu Zhanchun, Hao Wu
and Dr. Jinyang Liu for useful discussion. This work is supported by Major
Program of the Natural Science Foundation of China under Grant No. 22193032.
The author Tao Xu thanks the support of Peng Huanwu Foundation under
Contract No. 12447101.

\clearpage
 \centering
 \includegraphics[width=0.8\textwidth]{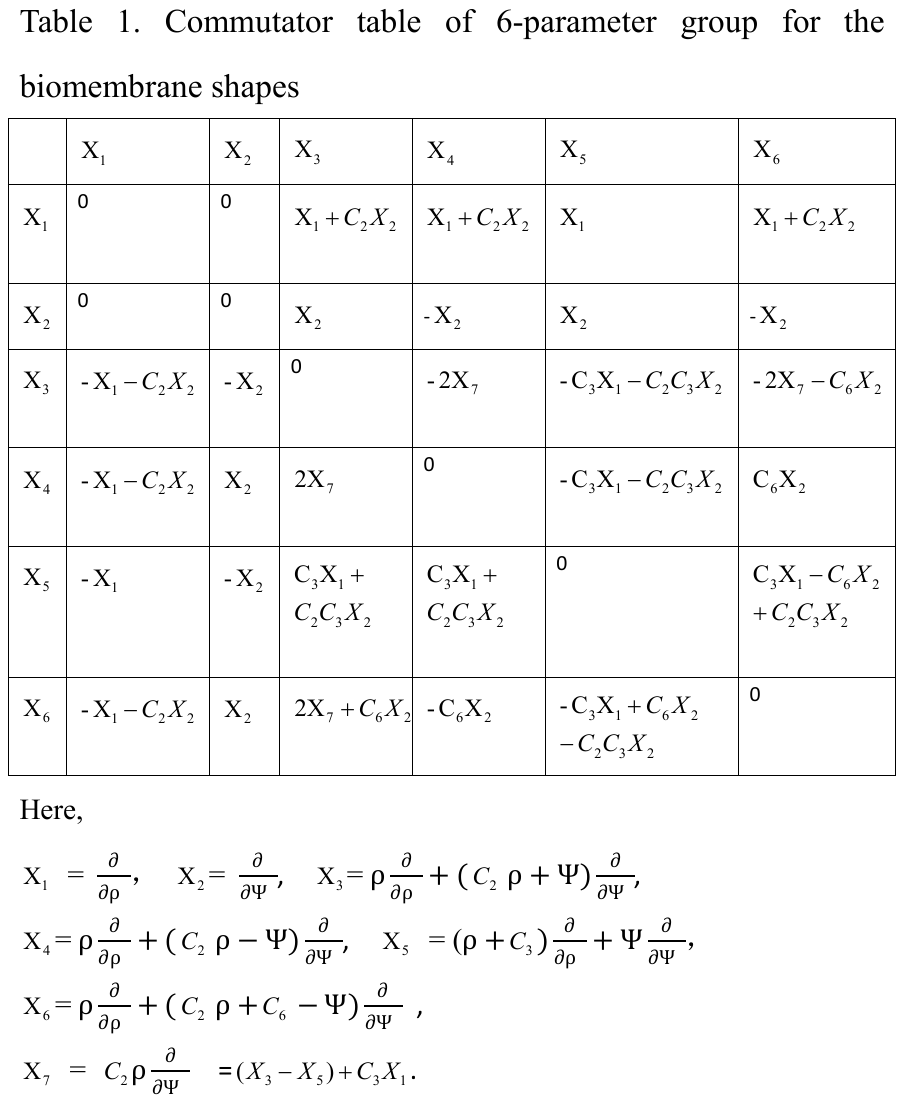}
 

\end{document}